\documentclass[10pt,twocolumn,twoside]{IEEEtran}

\usepackage{fixltx2e}
\usepackage{etex}
\usepackage[noadjust]{cite}
\usepackage[cmex10]{amsmath}
\usepackage{amssymb}
\usepackage{bm}
\usepackage{mathcomp}
\usepackage{dsfont}
\usepackage{ellipsis}
\usepackage{graphicx}
\usepackage[caption=false,font=footnotesize]{subfig}
\usepackage{enumerate}
\usepackage{flushend}		

\renewcommand{\log}[1]{\ensuremath{\text{log}\!\left(#1\right)}}		
\newcommand{\abs}[1]{\ensuremath{\lvert#1\rvert}}										
\newcommand{\norm}[1]{\ensuremath{\lVert#1\rVert}}									
\newcommand{\norminf}[1]{\ensuremath{\lVert#1\rVert_\infty}}				
\newcommand{\normone}[1]{\ensuremath{\lVert#1\rVert_1}}							
\newcommand{\vm}[1]{\ensuremath{\bm{\mathrm{#1}}}}									
\newcommand{\mc}[1]{\ensuremath{\mathcal{#1}}}											
\newcommand{\trans}{\ensuremath{\mathsf{T}}}												
\newcommand{\Z}{\ensuremath{\mathds{Z}}}														
\newcommand{\C}{\ensuremath{\mathbb{C}}}														
\newcommand{\sg}[2]{\ensuremath{g_{\mathrm{#1},#2}}}								
\newcommand{\sw}[1]{\ensuremath{w_{\mathrm{#1}}}}										
\newcommand{\swtilde}[1]{\ensuremath{\tilde{w}_{\mathrm{#1}}}}			
\newcommand{\spsi}[2]{\ensuremath{\psi_{\mathrm{#1},#2}}}						
\newcommand{\seta}[2]{\ensuremath{\eta_{\mathrm{#1},#2}}}						
\newcommand{\setatilde}[2]{\ensuremath{\tilde{\eta}_{\mathrm{#1},#2}}}   
\newcommand{\salpha}[2]{\ensuremath{\alpha_{\mathrm{#1},#2}}}				
\newcommand{\sbeta}[2]{\ensuremath{\beta_{\mathrm{#1},#2}}}					
\newcommand{\sgamma}[2]{\ensuremath{\gamma_{\mathrm{#1},#2}}}				

\begin{document}

\title{Linearization of Time-Varying Nonlinear Systems Using A Modified Linear Iterative Method}

\author{Matthias~Hotz,~\IEEEmembership{Student Member,~IEEE,}
        Christian~Vogel,~\IEEEmembership{Senior Member,~IEEE}%
\thanks{The research leading to these results has received funding from the FFG
Competence Headquarter program under the project number 835187.}%
\thanks{Matthias Hotz was with FTW, Austria. He is now with the Associate Institute for
Signal Processing, Technische Universit{\"a}t M{\"u}nchen, Germany (e-mail: hotz@ieee.org).}
\thanks{Christian Vogel is with FTW, Austria (e-mail: c.vogel@ieee.org).
The Austrian Competence Center FTW Forschungszentrum Telekommunikation Wien GmbH
is funded within the program COMET - Competence Centers for Excellent Technologies
by BMVIT, BMWFJ, and the City of Vienna. The COMET program is managed by the FFG.}}

\maketitle

\begin{abstract}
The linearization of nonlinear systems is an important digital
enhancement technique. In this paper, a real-time capable post- and pre-linearization
method for the widely applicable time-varying discrete-time Volterra
series is presented. To this end, an alternative view on the Volterra
series is established, which enables the utilization of certain modified
linear iterative methods for linearization. For one particular linear
iterative method, the Richardson iteration, the corresponding post- and pre-linearizers are discussed
in detail. It is motivated that the resulting algorithm can be regarded
as a generalization of some existing methods. Furthermore, a simply verifiable
condition for convergence is presented, which allows the straightforward
evaluation of applicability. The proposed method is demonstrated by means
of the linearization of a time-varying nonlinear amplifier, which highlights
its capability of linearizing significantly distorted signals, illustrates the
advantageous convergence behavior, and depicts its robustness against modeling errors.
\end{abstract}

\begin{IEEEkeywords}
Linearization, equalization, digital predistortion, nonlinear systems, time-varying systems,
Volterra series, iterative methods, Richardson iteration, $P$th-order inverse. 
\end{IEEEkeywords}

\IEEEpeerreviewmaketitle

\section{Introduction}
	\label{intro}

\IEEEPARstart{D}{igital} enhancement techniques became an effective
approach to improve the performance of analog systems due to rapid
advances in semiconductor technology~\cite{murmann2008}.
Linearization of nonlinear systems via real-time capable methods, as investigated in this paper,
is a particular digital enhancement technique. It is applied, e.g.,
to sensor linearization~\cite{zhuang2004}, amplifier predistortion~\cite{morgan2006},
channel equalization~\cite{benedetto1983}, and loudspeaker linearization~\cite{lashkari2006}.
The purpose of a linearizer is to compensate for the nonlinear behavior
of a nonlinear system, i.e., the nonlinear system in cascade with a
corresponding linearizer results in a defined linear behavior.
A special case thereof is equalization, where the targeted linear
behavior is the identity function.

Due to the lack of a unifying model for nonlinear systems, linearizers
are generally limited to a particular class of systems and, of course, linearization
is only possible if the nonlinear system preserves all information
(cf, e.g.,~\cite{geiger2011}). In this paper, nonlinearities
described by a \emph{time-varying discrete-time Volterra series} are considered. The
Volterra series is a widely used approximator for weakly nonlinear
systems~\cite{schetzen1980,rugh1981,mathews2000}. The objective of this paper is
the construction of a linearizer for a \emph{known} time-varying
discrete-time Volterra series. 
For the time-invariant Volterra series, the special case of equalization has already been
considered by Schetzen~\cite{schetzen1976} via the concept of a $P$th-order inverse.
A $P$th-order inverse is constructed by constraining the Volterra operators of the
overall system, i.e., the cascade of the nonlinear system and the $P$th-order inverse. The
constraint applies to the Volterra operators up to order $P$, whereas the operators
of higher order are arbitrary. Sarti and Pupolin~\cite{sarti1992} utilized
the fact that orders greater than $P$ of the overall system are not constrained
to derive a recursive synthesis scheme for a $P$th-order inverse that is less
complex compared to the $P$th-order inverse in~\cite{schetzen1976}. However,
the analysis of the existence and convergence of a $P$th-order inverse is
nontrivial and addressed, e.g., in~\cite{fang2001} for signals with finite energy.
An approach to linearization of the time-invariant Volterra series
is discussed by Nowak and Van Veen in~\cite{nowak1997}, where the
linearization problem is reformulated as a nonlinear fixed-point equation, which
is solved via successive approximation. They provide
an analysis of convergence with respect to a ``windowed $l^2$ norm,'' which, again,
turns out as a nontrivial task. Aschbacher et al.~\cite{aschbacher2004} (cf.~\cite{redfern1998} as well)
reduce the linearization problem for a time-invariant Volterra series to
the root-finding problem, which is solved using the Newton method. However,
for this iterative algorithm the crucial analysis of convergence is even more
involved and the corresponding conditions have not been reported yet.
As all these methods are limited to time-invariant nonlinear systems, it is worthwhile
to mention that for linear time-varying (LTV) systems equalization techniques have already
been introduced. They may be divided roughly into two classes~\cite{vogel2012},
explicitly designed correction filters, where the equalization problem is posed as a
filter design problem~\cite{johansson2006,johansson2008}, and iterative correction filters,
where the desired equalization result is approximated iteratively~\cite{vogel2009,
tsui2011b,soudan2012}.

\subsection{Contributions and Outline}
	\label{intro:contrib}

Real-world nonlinear systems often vary with time, e.g., due to
temperature variations or other environmental changes.
However, all existing methods for nonlinear systems reviewed
above consider only a \emph{time-invariant} Volterra series. Although
it is possible to extend the methods in~\cite{schetzen1976,sarti1992} and~\cite{nowak1997} to
the time-varying Volterra series~\cite{soudan2011}, they potentially become prohibitively complex
in computational terms. This stems from the fact that the $P$th-order inverse in~\cite{schetzen1976}
and~\cite{sarti1992} as well as the method in~\cite{nowak1997} require a (stable) inverse filter
for the first-order Volterra operator. This time-varying inverse filter is usually not known and,
in practice, has to be approximated using filter design techniques as discussed, e.g., in~\cite{vogel2012}.
Therefore, a change of the first-order Volterra operator implies the need for a computationally
costly filter design of its inverse.
Furthermore, the condition for convergence of all aforementioned methods is
either missing, very restrictive on the input signal, or rather complicate to
evaluate. Finally, it must be pointed out that in~\cite{soudan2011} a post-equalizer
for a time-varying Volterra series based on a nonlinear fixed point iteration is
discussed briefly, however, it completely lacks the critical analysis of convergence.
In this paper, the issues above are addressed via the following contributions:
\begin{enumerate}[a)]
	\item \emph{Alternative view on the Volterra series}: In Section~\ref{model}, an
		alternative description of the Volterra series
		is established, which provides a framework for the derivation of linearization
		methods based on certain modified linear iterative methods.
	\item \emph{Post- and pre-linearization}: Using this system model,
		a modification of the Richardson iteration is proposed in Section~\ref{mlim}, which permits its
		application for post- and pre-linearization of a time-varying
		discrete-time Volterra series as discussed in Sections~\ref{postlin} and~\ref{prelin}.
		The presented method is independent of the inverse of the first-order Volterra operator
		and, therefore, offers a computational advantage compared to the methods
		in~\cite{schetzen1976,sarti1992}, and~\cite{nowak1997} since the repeated and
		computationally costly inverse filter design is not necessary.
	\item \emph{Condition for convergence}: In Section~\ref{conv}, a sufficient condition
		for convergence is presented, which is particularly simple to evaluate and only requires
		a bounded input signal. Therewith, the applicability of the introduced linearization
		method is easily verified.
	\item \emph{Generalization}: In Section~\ref{relation}, it is shown that the
		proposed method is a generalization of the equalization
		method for LTV systems in~\cite{soudan2012}.
		Furthermore, it is motivated that the presented approach can be regarded as a generalization
		of the post-linearization method in~\cite{nowak1997} as well as the $P$th-order inverse.
\end{enumerate}
Section~\ref{results} presents simulation results, which demonstrate the
proposed method by means of the linearization of a time-varying nonlinear amplifier
and highlight its properties.
Finally, Section~\ref{conclusion} concludes the paper. The theory presented in this
paper requires two results for the time-varying discrete-time Volterra series which
have not been established yet and, therefore, are contributed via the appendix of
this paper:
\begin{enumerate}
	\item[e)] \emph{Properties of a time-varying discrete-time Volterra series:}
		In Appendix~\ref{bibo}, the conditions for the convergence
		of a time-varying discrete-time Volterra series are presented.
		Furthermore, in Appendix~\ref{errprop}, it is proven that a convergent time-varying discrete-time
		Volterra series is Lipschitz continuous.
\end{enumerate}

\subsection{Relation to Adaptive Nonlinear Equalization}

In practical application scenarios, the nonlinear system is usually not
known and, as a consequence, two approaches towards linearization emerge, i.e.,
(a) the direct identification of the linearizer, and
(b) the identification of the nonlinear system and construction
of the linearizer. For (a), adaptive nonlinear filters may be utilized, e.g.,~\cite{mathews1991}.
However, the identification is complicated by the model selection as
the structure of the linearizing system is usually not known.
In contrast, for (b) the nonlinear system is identified,
whose structure is often known, e.g., in terms of its circuit schematics,
topology, or physical properties. This simplifies the model selection and,
consequently, the identification, which motivates the utilization of the
approach in (b) that may use the construction of the linearizer
discussed in this paper. The identification of a Volterra series is discussed, e.g.,
in~\cite{mathews2000,glentis1999,nemeth2002,weng2006,giannakis2001}
and is not addressed in this paper. 

%
\begin{figure}[!t]
	\centering
	\includegraphics{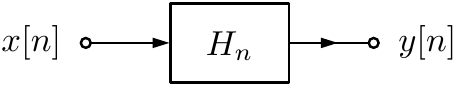}
	\caption{Volterra system $H_n$ with input signal $x[n]$ and output signal $y[n]$.}
	\label{fig:volterrasystem}
\end{figure}

%
\begin{figure*}[!t]
	\centering
	\includegraphics{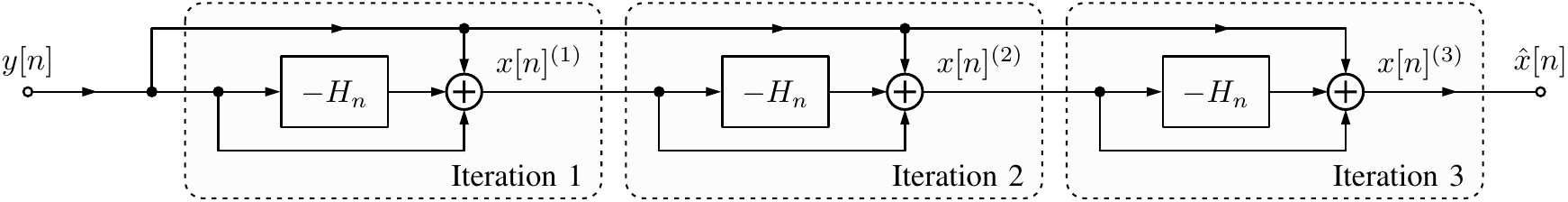}
	\caption{Richardson equalizer based on the Volterra system $H_n$
		with three iterations of~\eqref{eq:richeq} and the initialization in~\eqref{eq:richinit}.
		Due to the finite number of iterations, the reconstruction $\hat{x}[n]=x[n]^{(3)}$ is only an approximation
		of the desired reconstruction result $x[n]$.}
	\label{fig:richeq}
\end{figure*}
%

\section{System Model}
	\label{model}

The nonlinear system is modeled by a \emph{time-varying discrete-time Volterra series}, i.e.,
its complex-valued output sample $y[n]$ at time instant $n\in\Z$ is given
by~\cite{schetzen1980,rugh1981,mathews2000}
\begin{align}
	y[n] &= \sum_{p=1}^\infty \sum_{k_1,\dots,k_p\in\Z} h_{p,n}[k_1,\dots,k_p]\prod_{i=1}^p x[n-k_i]
	\label{eq:vsdef}
\end{align}
where $x[n]$ is the complex-valued input signal and $h_{p,n}$ are the
complex-valued \emph{time-varying Volterra
kernels}.\footnote{In its most general form, the Volterra series includes a term of order $0$,
i.e., a time-varying offset $h_{0,n}$. However, to simplify the discussion it is common
to require that the offset is compensated separately~\cite{mathews1991} and, therefore,
it is assumed that $h_{0,n}=0$.} Throughout this paper, it is assumed that the Volterra series
converges for the given input signal, cf. Appendix~\ref{bibo} for a discussion of convergence.
As a simplified representation, the time-varying Volterra series operator $H_n$ is defined
to describe the relation in~\eqref{eq:vsdef}, i.e.,
\begin{align}
	y[n] &= H_n\{x[n]\}
	\label{eq:hndef}
\end{align}
and the nonlinear system is referred to as the \emph{Volterra system} $H_n$ in the remainder
of the text, cf. Fig.~\ref{fig:volterrasystem}.
For the derivation of the linearization algorithm proposed in this paper,
a new view on the Volterra system is established.
To this end, the sum over $k_1$ in~\eqref{eq:vsdef} is evaluated as the outermost
sum and $x[n-k_1]$ is factored out, which permits the reformulation
\begin{align}
	y[n] = \sum_{k_1\in\Z} \sg{x}{n}[k_1] x[n-k_1]
	\label{eq:vsconv}
\end{align}
where
\begin{align}
	\begin{split}
	\sg{x}{n}[k_1] &= h_{1,n}[k_1] \\
		&+ \sum_{p=2}^\infty \sum_{k_2,\ldots,k_p\in\Z}
			 h_{p,n}[k_1,\ldots,k_p]\prod_{i=2}^p x[n-k_i]\;.
	\end{split}
	\label{eq:gxn}
\end{align}
The equivalent description in~\eqref{eq:vsconv} of the Volterra system
in~\eqref{eq:vsdef} resembles an LTV system with the time-varying impulse response
in~\eqref{eq:gxn}. However, the pretended impulse response $\sg{x}{n}[k_1]$ is not only time-varying
by means of a dependence of the coefficients on the time index $n$, which is denoted by
the subscript $n$, but depends on the input signal $x[n]$ as well, which is indicated by
the subscript $\mathrm{x}$ and symbolizes its \emph{nonlinear} nature.
The system description in~\eqref{eq:vsconv} can also be
cast in a matrix equation. Let $\C^\Z$ denote the space of bi-infinite
complex-valued sequences. The input vector $\vm{x}\in\C^\Z$ is defined as
\begin{align}
	\vm{x} = (\ldots, x[n+1], x[n], x[n-1], \ldots)^\trans
	\label{eq:xvec}
\end{align}
and comprises the samples of the input signal $x[n]$.
Analogously, the output vector $\vm{y}\in\C^\Z$ is defined as
\begin{align}
	\vm{y} = (\ldots, y[n+1], y[n], y[n-1], \ldots)^\trans
	\label{eq:yvec}
\end{align}
and comprises the samples of the output signal $y[n]$. Furthermore, an infinite
coefficient matrix $\vm{A}_{\vm{x}}$ is defined,
whose elements $(\vm{A}_{\vm{x}})_{ij}$ are given by
\begin{align}
	(\vm{A}_{\vm{x}})_{ij} = \sg{x}{i}[i-j]
	\label{eq:axdef}
\end{align}
in which $i,j\in\Z$ denote the row and column index, respectively, and
where the subscript $\vm{x}$ denotes the dependence on the input vector $\vm{x}$.
Therewith, the matrix equation
\begin{align}
	\vm{y} = \vm{A}_{\vm{x}}\vm{x}
	\label{eq:mtxeq}
\end{align}
constitutes an equivalent description of~\eqref{eq:vsconv} and, consequently,
of the Volterra system $H_n$ in~\eqref{eq:vsdef}.

\subsection{Problem Statement}
	\label{model:problem}

Consider the case of an equalizer that is connected to the output of
the Volterra system $H_n$ in Fig.~\ref{fig:volterrasystem}.
Then, equalization is the task of finding the input vector $\vm{x}$ given the
output vector $\vm{y}$ and the Volterra system $H_n$. 
With the previously introduced system model, the unknown input vector
$\vm{x}$ can be found by solving the ``system of linear equations'' in~\eqref{eq:mtxeq}.
Indeed, post- and pre-linearization can be recast
as a problem with such a structure, which is discussed later on.
However, the coefficient matrix $\vm{A}_{\vm{x}}$
depends on the solution $\vm{x}$ and, therefore, is \emph{unknown}.
Furthermore, if the resulting algorithm should be real-time capable and, thus,
reconstruct the signal sample by sample, it has to operate row by row with
respect to the matrix equation. Consequently, in order to solve the linearization problem
at hand, an algorithm to solve the ``system of linear equations'' in~\eqref{eq:mtxeq}
is required, which (a) \emph{operates row by row} and
(b) \emph{determines the coefficient matrix} $\vm{A}_{\vm{x}}$ alongside
the solution $\vm{x}$ by exploiting the structural knowledge.

\section{Modified Linear Iterative Method}
	\label{mlim}

There exist certain \emph{linear iterative methods}~\cite{kelley1995, isaacson1994, saad2003}
for solving systems of linear equations, whose iteration steps operate
row by row and, therefore, address problem (a) in Section~\ref{model:problem}.
These methods reformulate the problem of solving a system of linear
equations as a \emph{linear fixed-point problem}, which is solved using \emph{successive
approximation}~\cite{kelley1995}. A fixed-point equation comprises a function $\mc{T}$, where
the image of the solution $\vm{x}$ of the system of linear equations under $\mc{T}$
is $\vm{x}$~\cite{kelley1995}, i.e., $\vm{x} = \mc{T}(\vm{x})\:$.
There exist various approaches to rewrite a matrix equation of the form
in~\eqref{eq:mtxeq} as a fixed-point equation,
which eventually leads to different linear iterative methods~\cite{kelley1995,isaacson1994,saad2003}.
This paper focuses on the \emph{Richardson iteration},
but other linear iterative methods, e.g., the Jacobi and Gau\ss-Seidel iteration,
are applicable as well. Let $\vm{I}$ denote the identity matrix, then
adding $(\vm{I}-\vm{A}_{\vm{x}})\vm{x}$ to the left and right hand side
of~\eqref{eq:mtxeq} results in the fixed-point equation
\begin{align}
	\vm{x} = (\vm{I}-\vm{A}_{\vm{x}})\vm{x} + \vm{y}\;.
	\label{eq:richitervec}
\end{align}
If the fixed-point $\vm{x}$ is determined using successive approximation,
the \emph{Richardson iteration} is obtained~\cite{kelley1995}, i.e.,
\begin{align}
	\vm{x}^{(r+1)} = (\vm{I}-\vm{A}_{\vm{x}})\vm{x}^{(r)} + \vm{y}
	\label{eq:richiter}
\end{align}
in which $r$ is the iteration index.
Therefore, given an initial approximation $\vm{x}^{(0)}$ of the solution $\vm{x}$, this iteration
provides a sequence of approximations, which, under certain conditions, converges to the
fixed-point $\vm{x}$, i.e., $\lim_{r\rightarrow\infty} \vm{x}^{(r)} = \vm{x}\:$.

\subsection{Modified Richardson Iteration}

The Richardson iteration in~\eqref{eq:richiter} requires the knowledge of the
coefficient matrix $\vm{A}_{\vm{x}}$ and, consequently, cannot overcome problem
(b) in Section~\ref{model:problem}. It reconstructs the unknown input vector
$\vm{x}$ by iteratively improving an initial approximation $\vm{x}^{(0)}$ using the
output vector $\vm{y}$ and the \emph{unknown} coefficient matrix $\vm{A}_{\vm{x}}$.
However, in iteration $r+1$ the approximation $\vm{x}^{(r)}$ is already available and may
be used to approximate the coefficient matrix. To this end, in analogy to~\eqref{eq:axdef} the approximation
$\vm{A}_{\vm{x}^{(r)}}$ of $\vm{A}_{\vm{x}}$ based on $\vm{x}^{(r)}$ is defined in
terms of its elements $(\vm{A}_{\vm{x}^{(r)}})_{ij}$ in row $i$ and in column $j$ as
\begin{align}
	(\vm{A}_{\vm{x}^{(r)}})_{ij} = \sg{x^{(r)}}{i}[i-j]
	\label{eq:axrdef}
\end{align}
where $i,j\in\Z$ and
\begin{align}
	\begin{split}
	&\sg{x^{(r)}}{n}[k_1] = h_{1,n}[k_1] \\
		&\quad\ + \sum_{p=2}^\infty \sum_{k_2,\ldots,k_p\in\Z}
			h_{p,n}[k_1,\ldots,k_p]\prod_{i=2}^p x[n-k_i]^{(r)}\,.
	\end{split}
	\label{eq:gxrn}
\end{align}
Substitution of the coefficient matrix $\vm{A}_{\vm{x}}$ in the Richardson iteration
in~\eqref{eq:richiter} with the approximation $\vm{A}_{\vm{x}^{(r)}}$ yields
\begin{align}
	\vm{x}^{(r+1)} = (\vm{I}-\vm{A}_{\vm{x}^{(r)}})\vm{x}^{(r)} + \vm{y}\;.
	\label{eq:richeqvec}
\end{align}
This modified Richardson iteration does not only generate approximations of the input vector $\vm{x}$, but
also of the coefficient matrix $\vm{A}_{\vm{x}}$ and, consequently, overcomes problem (b)
in Section~\ref{model:problem} as well. Indeed, it provides a solution to a
significant class of systems as shown by the condition for convergence discussed
in Section~\ref{conv} later on.

\subsection{Richardson Equalizer}
	\label{mlim:richeq}

%
\begin{figure}[!t]
	\centering
	\includegraphics{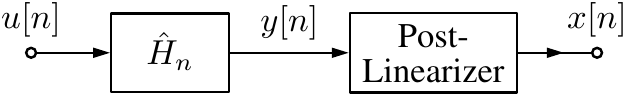}
	\caption{The post-linearizer for the Volterra system $\hat{H}_n$
		generates an output signal $x[n]$ that corresponds to the response of the
		LTI filter $L$ to the input signal $u[n]$, cf.~\eqref{eq:xludef}.}
	\label{fig:hnpostlin:a}
\end{figure}
\begin{figure}[!t]
	\centering
	\includegraphics{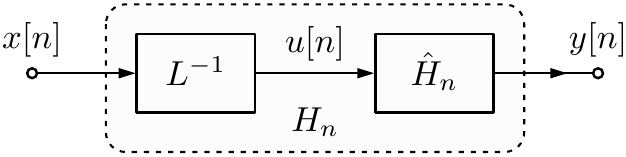}
	\caption{Stimulation of the augmented Volterra system $H_n$ with the signal $x[n]$
		results in the same signal $y[n]$ as in Fig.~\ref{fig:hnpostlin:a} if the LTI
		filter $L$ that relates $x[n]$ and $u[n]$ possesses a stable inverse filter $L^{-1}$.}
	\label{fig:postlinaltsys}
\end{figure}

The modified Richardson iteration in~\eqref{eq:richeqvec} is a matrix equation, but
for a real-time capable algorithm, the iteration needs to be sample-based. Therefore,
consider the evaluation of~\eqref{eq:richeqvec} row by row. Using~\eqref{eq:xvec},
\eqref{eq:yvec}, and~\eqref{eq:axrdef}, this results in
\begin{align*}
	x[n]^{(r+1)} = x[n]^{(r)} + y[n] - \sum_{k_1\in\Z} \sg{x^{(r)}}{n}[k_1] x[n-k_1]^{(r)}\;.
\end{align*}
A comparison of the convolution of $\sg{x^{(r)}}{n}[k_1]$ with $x[n]^{(r)}$ to~\eqref{eq:vsconv}
reveals that it equals the response of the Volterra system $H_n$ to the
input signal $x[n]^{(r)}$. Hence, a real-time capable algorithm based on the modified Richardson iteration,
for convenience called \emph{Richardson equalizer} in the remainder, is described by the
iteration
\begin{align}
	x[n]^{(r+1)} = x[n]^{(r)} + y[n] - H_n\{x[n]^{(r)}\}\;.
	\label{eq:richeq}
\end{align}
If $H_n$ is causal,~\eqref{eq:richeq} is indeed realizable as a sample-based iteration,
i.e., the reconstruction $x[n]^{(r+1)}$ of the sample $x[n]$ depends only on previous
reconstructions $x[k]^{(r)}$, where $k \leq n$. 
While the approximation $x[n]^{(r)}$ will equal the desired input sample $x[n]$
if $r\rightarrow\infty$ and if the iteration converges,
a practical system can, of course, only implement a finite number of iterations.
This is not a limitation per se as the number of iterations can be
chosen so that the accuracy of the approximation suffices for the specific application.
However, this statement assumes that the approximation improves in every iteration or,
in other words, the error with respect to the solution decreases in every iteration.
This iterative reduction of the error is indeed ensured by the conditions for
convergence discussed in Section~\ref{conv}.
Still, the finite number of iterations introduces another issue, i.e.,
the initialization $x[n]^{(0)}$ influences the approximation accuracy.
While the initialization may be chosen arbitrarily, e.g., $x[n]^{(0)} = 0$,
it should be as close to the solution $x[n]$ as possible to improve the approximation result.
In Section~\ref{conv}, it is shown that the iteration in~\eqref{eq:richeq} converges
for \emph{moderately} nonlinear systems.
Under these circumstances, the output is a rough approximation of the input and
it turns out to be advantageous to use the initialization
\begin{align}
	x[n]^{(0)} = y[n]\;.
	\label{eq:richinit}
\end{align}
Concluding, a Richardson equalizer based on three iterations of~\eqref{eq:richeq}
with the initialization in~\eqref{eq:richinit} is depicted in Fig.~\ref{fig:richeq}.

%
\begin{figure}[!t]
	\centering
	\includegraphics{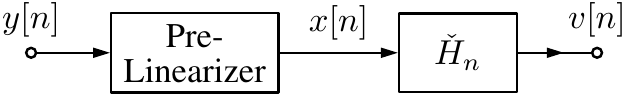}
	\caption{The pre-linearizer for the Volterra system $\check{H}_n$
		generates a signal $x[n]$ to which $\check{H}_n$ responds with a signal $v[n]$
		that corresponds to the response of the
		LTI filter $L$ to the input signal $y[n]$, cf.~\eqref{eq:vlydef}.}
	\label{fig:hnprelin:a}
\end{figure}
\begin{figure}[!t]
	\centering
	\includegraphics{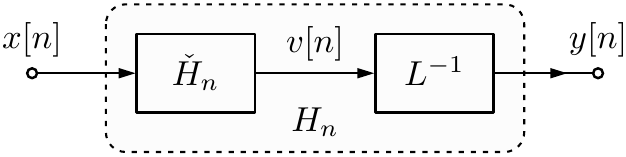}
	\caption{Stimulation of the augmented Volterra system $H_n$ with the signal $x[n]$
		results in the same signal $y[n]$ as in Fig.~\ref{fig:hnprelin:a} if the LTI
		filter $L$ that relates $v[n]$ and $y[n]$ possesses a stable inverse filter $L^{-1}$.}
	\label{fig:prelinaltsys}
\end{figure}
%

\section{Post-Linearization}
	\label{postlin}

Linearization is the problem of correcting the nonlinear behavior of a
given system to a defined linear behavior by cascading it with another
system, where the latter system is called
\emph{linearizer}. As the cascade of two nonlinear systems,
in general, exhibits a different behavior depending on the ordering of the
systems, two configurations arise, i.e., post- and pre-linearization.
In the case of post-linearization of a Volterra system $\hat{H}_n$, the
linearizer is connected to the output of $\hat{H}_n$ as depicted in
Fig.~\ref{fig:hnpostlin:a} and termed \emph{post-linearizer}
to distinguish it from the pre-linearizer discussed in Section~\ref{prelin}.
The desired behavior of the cascade is described by the linear time-invariant
(LTI) filter $L$, i.e., the output signal $x[n]$ of the cascade is given by
\begin{align}
	x[n]=L\{u[n]\}
	\label{eq:xludef}
\end{align}
in which $u[n]$ is the input signal of the cascade. Consequently, the task of the
post-linearizer is to reconstruct the signal $x[n]$ by observation of the signal $y[n]$ and
knowledge of the Volterra system $\hat{H}_n$. Given that the LTI filter $L$ is
minimum-phase and, thus, possesses a stable inverse filter $L^{-1}$, i.e.,
\begin{align}
	u[n] = L^{-1}\{x[n]\}
	\label{eq:linvdef}
\end{align}
the signal $y[n]$ in Fig.~\ref{fig:hnpostlin:a} can be regarded as the response
of an augmented Volterra system $H_n$ to the input signal $x[n]$.
To this end, consider that the Volterra system $H_n$ is the cascade of the
inverse filter $L^{-1}$ and the Volterra system $\hat{H}_n$, i.e.,
\begin{align}
	H_n = L^{-1} \circ \hat{H}_n
	\label{eq:hnbreve}
\end{align}
in which $\circ$ denotes cascade connection. Then, the signal $y[n]$
can be described as
\begin{align*}
	y[n] = \hat{H}_n\{u[n]\}
	     = \hat{H}_n\{L^{-1}\{x[n]\}\}
	     = H_n\{x[n]\}
\end{align*}
which is illustrated in Fig.~\ref{fig:postlinaltsys}.
This corresponds to the relation in~\eqref{eq:hndef},
where the Richardson equalizer introduced in Section~\ref{mlim} provides
the means to reconstruct the signal $x[n]$. Consequently, the Richardson
equalizer described by the iteration~\eqref{eq:richeq} based on the Volterra
system $H_n$ in~\eqref{eq:hnbreve} with the initialization in~\eqref{eq:richinit}
constitutes a post-linearizer for the Volterra system $\hat{H}_n$.

\section{Pre-Linearization}
	\label{prelin}

In the case of pre-linearization of a Volterra system $\check{H}_n$,
often called \emph{digital predistortion} as well, the
linearizer is connected to the input of $\check{H}_n$ as depicted in
Fig.~\ref{fig:hnprelin:a} and termed \emph{pre-linearizer}.
The desired behavior of the cascade is described by the LTI filter $L$,
i.e., the output signal $v[n]$ of the cascade is given by
\begin{align}
	v[n]=L\{y[n]\}
	\label{eq:vlydef}
\end{align}
in which $y[n]$ is the input signal of the cascade. Consequently, the task of the
pre-linearizer is to reconstruct the signal $x[n]$ by observation of the signal $y[n]$ and
knowledge of the Volterra system $\check{H}_n$ so that $x[n]$ filtered by $\check{H}_n$
results in the desired output signal $v[n]$ in~\eqref{eq:vlydef}. Analogous to
Section~\ref{postlin}, the signal $y[n]$ in Fig.~\ref{fig:hnprelin:a} can be regarded
as the response of an augmented Volterra system $H_n$ to the input signal $x[n]$
if the LTI filter $L$ possesses a stable inverse filter $L^{-1}$.
To this end, consider that the Volterra system $H_n$ is the cascade of
the Volterra system $\check{H}_n$ and the inverse filter $L^{-1}$, i.e.,
\begin{align}
	H_n =  \check{H}_n \circ L^{-1}\;.
	\label{eq:hnring}
\end{align}
Then, the signal $y[n]$ can be described as
\begin{align*}
	y[n] = L^{-1}\{v[n]\}
	     = L^{-1}\{\check{H}_n\{x[n]\}\}
	     = H_n\{x[n]\}
\end{align*}
which is illustrated in Fig.~\ref{fig:prelinaltsys}.
Again, it can be observed that this corresponds to the
relation in~\eqref{eq:hndef} and, thus, the Richardson equalizer provides
the means to synthesize the signal $x[n]$. Consequently, the Richardson
equalizer described by the iteration~\eqref{eq:richeq} based on the Volterra
system $H_n$ in~\eqref{eq:hnring} with the initialization in~\eqref{eq:richinit}
constitutes a pre-linearizer for the Volterra system $\check{H}_n$.

It is important to recognize that there is a difference in the approximation
mechanism between post- and pre-linearization if the Richardson equalizer
is utilized with a \emph{finite} number of iterations. With post-linearization, the
approximation generated by the Richardson iteration appears directly at the
output, i.e., any decrease in the approximation error is directly visible.
In the case of pre-linearization, the approximation generated by the Richardson
iteration traverses the Volterra system $\check{H}_n$ before appearing at the output
of the cascade, i.e., the approximation is subject to a \emph{nonlinear} filtering
operation. However, as shown in Appendix~\ref{errprop}, a time-varying discrete-time
Volterra series is \emph{Lipschitz continuous} if it converges.
This implies that if the approximation error at the input decreases, the upper
bound on the approximation error at the output decreases as well. In other words,
an improvement in approximation accuracy at the input results in an improvement of
the worst-case approximation accuracy at the output and, consequently, the
application of the pre-linearizer is indeed appropriate.

\section{Conditions for Convergence}
	\label{conv}

The application of the Richardson equalizer or any
other iterative method is only reasonable if the iteration converges to the
solution. In order to study the conditions under which convergence of the
Richardson equalizer can be guaranteed, the error $\vm{e}^{(r)}$ in
iteration $r$ is defined as
\begin{align}
	\vm{e}^{(r)} = \vm{x} - \vm{x}^{(r)}
	\label{eq:er1def}
\end{align}
in which \vm{x} is the solution that satisfies the fixed-point equation
in~\eqref{eq:richitervec} and $\vm{x}^{(r)}$ is the approximation in
iteration $r$ of the Richardson equalizer~\eqref{eq:richeqvec} in matrix notation.
The subtraction of~\eqref{eq:richeqvec} from~\eqref{eq:richitervec} and
utilization of~\eqref{eq:er1def} results in
\begin{align}
	\vm{e}^{(r+1)}
		&= (\vm{I}-\vm{A}_{\vm{x}})\vm{x} - (\vm{I}-\vm{A}_{\vm{x}^{(r)}})\vm{x}^{(r)} \notag\\
		&= (\vm{I}-\vm{A}_{\vm{x}})(\vm{x}^{(r)}+\vm{e}^{(r)})
				- (\vm{I}-\vm{A}_{\vm{x}^{(r)}})\vm{x}^{(r)} \notag\\
		&= (\vm{I} - \vm{A}_{\vm{x}})\vm{e}^{(r)}
				+ (\vm{A}_{\vm{x}^{(r)}} - \vm{A}_{\vm{x}})\vm{x}^{(r)}
	\label{eq:er1eq}
\end{align}
which depicts the influence of the error $\vm{e}^{(r)}$ in the previous
iteration and the approximation error $\vm{A}_{\vm{x}^{(r)}} - \vm{A}_{\vm{x}}$
of the coefficient matrix. The Richardson equalizer converges to $\vm{x}$ if
the error decays to zero, i.e., $\lim_{r\rightarrow\infty} \vm{e}^{(r)} = \vm{0}\:$.
An even more restrictive requirement is
\begin{align}
	\norminf{\vm{e}^{(r+1)}} < \norminf{\vm{e}^{(r)}}
	\label{eq:ermondec}
\end{align}
which has to hold for all iterations $r\geq 0$.
In~\eqref{eq:ermondec}, $\norminf{\cdot}$ denotes the supremum norm~\cite{rudin1964}, i.e.,
it requires the supremum of the error signal to be \emph{strictly} monotonically
decreasing with respect to the iteration index $r$.\footnote{Note
that any valid norm may be used in~\eqref{eq:ermondec} and that the choice has an impact on the derivation
and the resulting condition for convergence. Due to its beneficial structure,
the supremum norm is employed.} In this case,
the approximation error needs to decrease in every iteration, which corresponds to
the requirement on the Richardson equalizer for a finite number of iterations
identified in Section~\ref{mlim:richeq}. Using~\eqref{eq:er1eq},
it is shown in Appendix~\ref{ccderi1} that if the function
\begin{align}
	\spsi{x}{n} = \sum_{k_1\in\Z} \abs{\delta[k_1]-h_{1,n}[k_1]}
		+ \sum_{p=2}^\infty \normone{h_{p,n}}\cdot \sw{x}(p)
	\label{eq:richpsi}
\end{align}
of a Volterra system $H_n$ satisfies the \emph{condition for convergence}
\begin{align}
	\sup_{n\in\Z} \spsi{x}{n} < 1
	\label{eq:richcondconv}
\end{align}
then \eqref{eq:ermondec} holds for the Richardson equalizer in~\eqref{eq:richeq}
with the initialization in~\eqref{eq:richinit}.
In~\eqref{eq:richpsi}, $\delta[k_1]$ denotes the unit impulse sequence, i.e.,
\begin{align}
	\delta[k_1] =
		\left\{
			\begin{array}{ll}
				1\,, & \text{if}\ k_1 = 0\\
				0\,, & \text{if}\ k_1 \neq 0\;.
			\end{array}
		\right.
	\label{eq:deltadef}
\end{align}
$\normone{h_{p,n}}$ is defined as the sum of the absolute coefficients of the
$p$th-order Volterra kernel at time instant $n$, i.e.,
\begin{align}
	\normone{h_{p,n}} = \sum_{k_1,\ldots,k_p\in\Z} \abs{h_{p,n}[k_1,\ldots,k_p]}
	\label{eq:hpnnorm}
\end{align}
and the weighting factor $\sw{x}(p)$ is given by
\begin{align}
	\sw{x}(p) = (2^p-1) \norminf{\vm{x}}^{p-1}\;.
	\label{eq:wdef}
\end{align}
For practical systems, which operate only for a finite time,
the condition in~\eqref{eq:richcondconv} is particularly
simple to verify as it suffices to ensure that $\spsi{x}{n}<1$
holds at \emph{every} time instant $n$. This is simply a threshold on a weighted
sum of the absolute kernel coefficients, where the weights
depend on the input amplitude range which is usually known.

It follows from the condition in~\eqref{eq:richcondconv}
that the rate of time-variance of the Volterra system
has no impact on whether the Richardson equalizer converges as long as
$\spsi{x}{n}$ consistently remains below the threshold. Furthermore, it is worthwhile to mention that the coefficient $h_{1,n}[0]$,
i.e., the coefficient of the first-order Volterra kernel
at time lag zero, is of particular significance since only its
difference from one contributes to the sum in~\eqref{eq:richpsi}.
Considering that the threshold on $\spsi{x}{n}$ in~\eqref{eq:richcondconv} is one,
this implies that the coefficient $h_{1,n}[0]$ is restricted to the open
interval $(0,2)$ and, in general, it must be dominant, i.e.,
all other coefficients must be small compared to
$h_{1,n}[0]$. However, by appropriately delaying signals and
matching time indices, this restriction may be loosened
to some arbitrary coefficient of the first-order
Volterra kernel, instead of being mandatory for the coefficient at time lag $0$.
As the corresponding structural modifications equal those for the method
in~\cite{soudan2012}, a detailed discussion thereof is omitted here
(see also Section~\ref{relation:ltveq}).

\subsection{Remarks to the Condition for Linearization}

If the Richardson equalizer is utilized for post- and pre-linearization,
the Volterra system $H_n$ is the cascade of an LTI filter and a Volterra
system as given by~\eqref{eq:hnbreve} and~\eqref{eq:hnring}.
In order to discuss the implications thereof on the condition for convergence,
let the inverse filter $L^{-1}$ of the minimum-phase LTI filter $L$
be characterized by the impulse response $q[n]$, i.e.,
\begin{align}
	L^{-1}\{x[n]\} = \sum_{l\in\Z} q[l] x[n-l]\;.
	\label{eq:linvdefconv}
\end{align}
As shown in Appendix~\ref{linkernels}, the kernels of the Volterra
system $H_n$ in~\eqref{eq:hnbreve} for post-linearization are given by
\begin{align}
	\begin{split}
	&h_{p,n}[k_1,\dots,k_p] \\
		&\quad\quad= \sum_{l_1,\dots,l_p\in\Z}  \hat{h}_{p,n}[k_1-l_1,\dots,k_p-l_p] \prod_{j=1}^p q[l_j]
	\end{split}
	\label{eq:hpnpostlindef}
\end{align}
and the kernels of the Volterra system $H_n$ in~\eqref{eq:hnring}
for pre-linearization are given by
\begin{align}
	h_{p,n}[k_1,\dots,k_p]
		= \sum_{l\in\Z} \check{h}_{p,n-l}[k_1-l,\dots,k_p-l] q[l]\;.
	\label{eq:hpnprelindef}
\end{align}
Those results show that the rather restrictive condition for
convergence in~\eqref{eq:richcondconv} is mitigated as it applies
to the Volterra systems $\hat{H}_n$ and $\check{H}_n$ only \emph{relative}
to the LTI filter $L$. To exemplify this, consider $\hat{H}_n$ and
$\check{H}_n$ to model a nonlinear amplifier of gain $K>0$, where the
desired behavior $L$ is an ideal amplifier with gain $K$. Thus,
$L^{-1}$ is characterized by
\begin{align}
	q[n] =
		\left\{
			\begin{array}{ll}
				1/K\,, & \text{if}\ n = 0\\
				0\,,   & \text{if}\ n \neq 0\;.
			\end{array}
		\right.
	\label{eq:linvqdef}
\end{align}
In case of pre-linearization, it follows from~\eqref{eq:hpnprelindef}
that the kernels $h_{p,n}$ equal the kernels $\check{h}_{p,n}$ weighted by $1/K$.
Therefore, the coefficients are weighted so that only the nonlinearity
relative to the linear gain has impact on the condition for convergence.
In case of post-linearization, it follows from~\eqref{eq:hpnpostlindef} that the
kernels $h_{p,n}$ equal the kernels $\hat{h}_{p,n}$ weighted by $1/K^p$.
However, in this setting the linearizer operates on the amplified signal,
cf. Fig.~\ref{fig:hnpostlin:a}. To investigate the implications in terms
of the unamplified signal $u[n]$, it is recognized from~\eqref{eq:xludef}
that $\norminf{\vm{x}}=K\norminf{\vm{u}}$. It can be seen from the
weighting factor in~\eqref{eq:wdef} that this amplification results in
an additional factor $K^{p-1}$ for $p\geq 2$. Therefore, the weighting
of the kernels $\hat{h}_{p,n}$ does not only relate them to $L$ by
weighting with $1/K$, but also accounts for the change in signal amplitude by
including the factor $1/K^{p-1}$.

\section{Relation to Other Methods}
	\label{relation}

%
\begin{figure}[!t]
	\centering
	\includegraphics{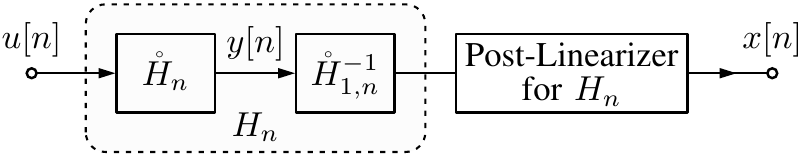}
	\caption{
		Equivalent implementation of the post-linearization method in~\cite{nowak1997}
		and the $P$th-order inverse in~\cite{kafka2002} for the Volterra system
		$\mathring{H}_n$ using the post-linearizer in Section~\ref{postlin} with
		$L$ set to the identity function.}
	\label{fig:nowakvsrich}
\end{figure}

\subsection{Equalization of Linear Weakly Time-Varying Systems}
	\label{relation:ltveq}

Soudan and Vogel~\cite{soudan2012} proposed an equalizer for linear weakly time-varying systems
which is based on the Richardson iteration. The method proposed in this paper
can be regarded as the generalization of the method in~\cite{soudan2012} from linear
to nonlinear systems and from equalization to linearization. In particular, if the
Volterra system $H_n$ comprises only a linear (first-order) kernel, the
Richardson equalizer equals the iteration in~\cite{soudan2012}. Furthermore,
for a linear system the condition for convergence in~\eqref{eq:richcondconv} reduces to the
criterion provided in~\cite{soudan2012}.

\subsection{Nonlinear Iterative Methods}

Instead of applying modified linear iterative methods to~\eqref{eq:mtxeq},
it is possible to directly formulate a nonlinear fixed-point equation
based on~\eqref{eq:vsdef} and solve it via successive approximation as
presented by Nowak and Van Veen~\cite{nowak1997}. However, let the
first-order Volterra operator $\mathring{H}_{1,n}$
of the Volterra system $\mathring{H}_n$ be defined as
\begin{align*}
	\mathring{H}_{1,n}\{x[n]\} = \sum_{k_1\in\Z} \mathring{h}_{1,n}[k_1] x[n-k_1]
\end{align*}
and possess an inverse $\mathring{H}^{-1}_{1,n}$,
which is a fundamental assumption in~\cite{nowak1997}.
Then, the post-linearizer in~\cite{nowak1997} for $\mathring{H}_n$, which is realized as a
post-equalizer followed by an LTI filter, equals
the post-linearizer in Section~\ref{postlin} followed by the same LTI filter,
if the latter linearizer is based on the Volterra system
\begin{align}
	H_n = \mathring{H}_n \circ \mathring{H}^{-1}_{1,n}
	\label{eq:hncheck}
\end{align}
and $L$ is set to the identity function, see Fig.~\ref{fig:nowakvsrich}.
In fact, it equals the extension of the post-linearizer in~\cite{nowak1997}
to \emph{time-varying} systems and also illustrates the dependence on the inverse
$\mathring{H}^{-1}_{1,n}$ considered in Section~\ref{intro:contrib}.
Consequently, the presented method may be regarded as a generalization
of the post-linearizer in~\cite{nowak1997}, as the latter amounts to the application
of the proposed post-linearizer to the augmented Volterra system $H_n$ in~\eqref{eq:hncheck}.

\subsection{$P$th-Order Inverse}

Due to the fact that the definition of a $P$th-order inverse does
not constrain the Volterra kernels of order greater than $P$ of the
overall system, different realizations exist~\cite{schetzen1976,sarti1992,kafka2002}.
If the post-linearizer in Section~\ref{postlin}, with $L$ set to the identity
function, is applied to $H_n$ in~\eqref{eq:hncheck} using the initialization
\begin{align}
	x[n]^{(0)} = H^{-1}_{1,n}\{y[n]\}
	\label{eq:poiequiinit}
\end{align}
the resulting iteration equals the extension of the recursive synthesis technique for a
$P$th-order inverse in~\cite[ch.~5.2.3]{kafka2002} to \emph{time-varying} systems,
cf. Fig~\ref{fig:nowakvsrich}.\footnote{The $P$th-order inverse
in~\cite{kafka2002} is specified by the recursive scheme (5.24) therein. Adding
$u_{p-1}[n] - H_1^{-1}\{ H_1\{ u_{p-1}[n] \} \} = 0$ to this equation, utilizing the
linearity of $H_1^{-1}$, and recognizing that $u_1[n] = H_1^{-1}\{y[n]\}$ leads to
$u_p[n] = u_{p-1}[n] + H_1^{-1}\{y[n]\} - H_1^{-1}\{ H\{ u_{p-1}[n] \} \}$,
in which $H\{ u_{p-1}[n] \} = H_1\{ u_{p-1}[n]\} + H_\text{NL}\{ u_{p-1}[n] \}$. This
recursive scheme corresponds to the iteration implemented by the post-linearizer in
Fig.~\ref{fig:nowakvsrich}.}
That is, the reconstruction after $r$ iterations corresponds to the reconstruction
of the ${(r+1)}$th-order inverse. Consequently, the
presented method may be regarded as a generalization of the $P$th-order inverse
as well, as the latter constitutes a particular application of the presented post-linearizer.

\section{Simulation Results}
	\label{results}

In the following, the post- and pre-linearization methods introduced in this
paper are demonstrated by means of the linearization of a
nonlinear amplifier with time-varying gain and dynamic saturation.
This amplifier is modeled by the Volterra system $\hat{H}_n$ comprising the kernels
\begin{align*}
	\hat{h}_{1,n}(k_1) &= \kappa_n \vm{c}_1(k_1)\\
	\hat{h}_{3,n}(k_1,k_2,k_3) &= \kappa_n \vm{c}_3(k_1) \vm{c}_3(k_2) \vm{c}_3(k_3)\\
	\hat{h}_{5,n}(k_1,k_2,k_3,k_4,k_5) &=
		\kappa_n \vm{c}_5(k_1) \vm{c}_5(k_2) \vm{c}_5(k_3) \vm{c}_5(k_4) \vm{c}_5(k_5)
\end{align*}
in which the coefficient vectors $\vm{c}_1$, $\vm{c}_3$, and $\vm{c}_5$
with zero-based element indexing are given by
\begin{align}
\begin{split}
	\vm{c}_1 &= ( 1.00,  0.03,  0.015)\\
	\vm{c}_3 &= (-0.38, -0.07, -0.03)\\
	\vm{c}_5 &= (-0.27, -0.06)
\end{split}
	\label{eq:coeffnlamp}
\end{align}
and the time-varying gain $\kappa_n$ is defined as
\begin{align*}
	\kappa_n = K\cdot [1 + 0.03\cos(4\pi n/N)]\;.
\end{align*}
In the latter, $K=50$ is the fundamental gain of the amplifier and
$N=500$ denotes the number of samples used for the simulation. For the sake
of consistent notation, an equivalent Volterra system $\check{H}_n=\hat{H}_n$
is defined for pre-linearization. The desired behavior of the amplifier is
an ideal gain of factor $K$, i.e., the LTI filter $L$ implements $L\{x[n]\} = K x[n]\:$.
Consequently, its inverse $L^{-1}$ is characterized by the impulse
response $q[n]$ in~\eqref{eq:linvqdef}. The input to the amplifier shall
be bounded by $B$ and, therefore, it follows from Fig.~\ref{fig:hnpostlin:a}
and~\eqref{eq:xludef} that for post-linearization
\begin{align}
	\norminf{\vm{x}} = K\norminf{\vm{u}} = KB
	\label{eq:simpostxinf}
\end{align}
and from Fig.~\ref{fig:hnprelin:a} that for pre-linearization
\begin{align}
	\norminf{\vm{x}} = B\;.
	\label{eq:simprexinf}
\end{align}
The input signal to the nonlinear amplifier is the modulated sine wave
\begin{align}
	s[n] = B\sin(2\pi n/N)\sin(38\pi n/N)\;.
	\label{eq:sndef}
\end{align}
Consequently, the desired output
signal is $K s[n]$. To achieve this output, the input is set to $u[n]=s[n]$
for post-linearization in Fig.~\ref{fig:hnpostlin:a} and to $y[n]=s[n]$ for
pre-linearization in Fig.~\ref{fig:hnprelin:a}.
Depending on the bound $B$ on the input of the nonlinear amplifier,
the distortion of the output signal varies and, in the following,
the linearization is studied for mildly, moderately, and severely
distorted output signals. Subsequently, the section concludes with
an investigation of the influence of modeling errors.

\subsection{Mild Distortion}
	\label{results:mild}

%
\begin{figure}[!t]\centering%
\includegraphics{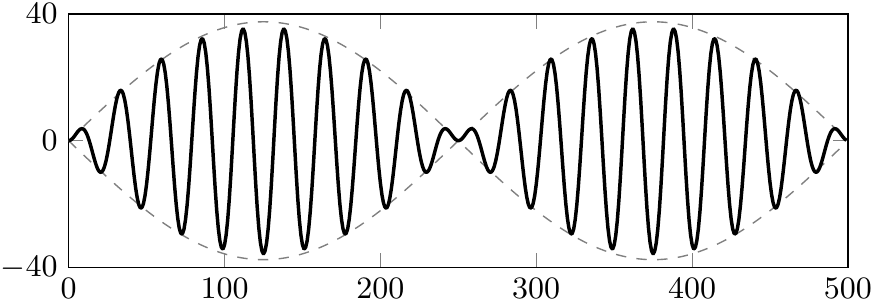}%
\caption{The solid line depicts the output signal of the nonlinear
	amplifier for $B=0.75$ when stimulated with $s[n]$ in~\eqref{eq:sndef}. To support the
	visual evaluation of the time-varying gain and dynamic saturation,
	the envelope of the desired output signal is shown as a dashed line.}%
\label{fig:inoutcmpb075}%
\end{figure}%
%
\begin{figure}[!t]\centering%
\includegraphics{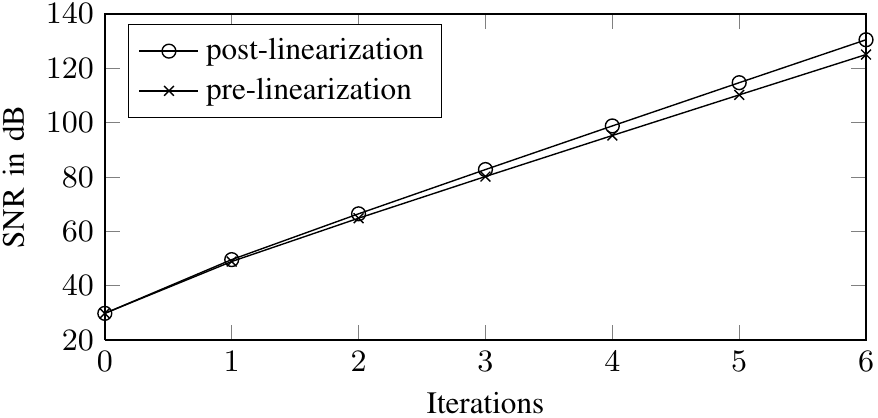}%
\caption{SNR after post- and pre-linearization of the nonlinear amplifier with $B=0.75$
	with respect to the number of iterations performed in the Richardson equalizer.
	For $0$ iterations, the SNR relates to the output signal without linearization.}%
\label{fig:snrb075}%
\end{figure}

For $B=0.75$ the output signal of the nonlinear amplifier is only
mildly distorted as shown in Fig.~\ref{fig:inoutcmpb075}. The applicability
of the post- and pre-linearization methods is verified
using~\eqref{eq:simpostxinf} and~\eqref{eq:simprexinf} and the definition
of $H_n$ in~\eqref{eq:hpnpostlindef} and~\eqref{eq:hpnprelindef}
in~\eqref{eq:richpsi}, respectively, to determine $\spsi{x}{n}$.
The maxima of $\spsi{x}{n}$ are at $0.5644$, which is significantly less
than one, and thus the condition in~\eqref{eq:richcondconv} guarantees convergence.
The linearization performance is measured with the signal-to-noise ratio (SNR)
\begin{align*}
	\text{SNR}\{x[n]\} = 10\cdot\log{\frac{\sum_{n=0}^{N-1}
		\abs{K s[n]}^2}{\sum_{n=0}^{N-1} \abs{K s[n] - x[n]}^2}}
\end{align*}
which is a logarithmic measure for the deviation from the desired output signal $K s[n]$. 
Post- and pre-linearization is performed with the Richardson equalizer in~\eqref{eq:richeq}
using $H_n$ in~\eqref{eq:hnbreve} and~\eqref{eq:hnring}, respectively, and the initialization
in~\eqref{eq:richinit}. In Fig.~\ref{fig:snrb075}, the linearization performance is depicted
in terms of SNR with respect to the number of iterations employed in the Richardson
equalizer. It can be observed that both linearizers converge very fast. The improvement in
SNR per iteration is significant and it increases approximately linear with the number of
iterations. The performance for pre-linearization is somewhat inferior to that of
post-linearization, which is primarily a consequence of the nonlinear filtering of the
approximation as discussed in Section~\ref{prelin}.

\subsection{Moderate Distortion}
	\label{results:moderate}

%
\begin{figure}[!t]\centering%
\includegraphics{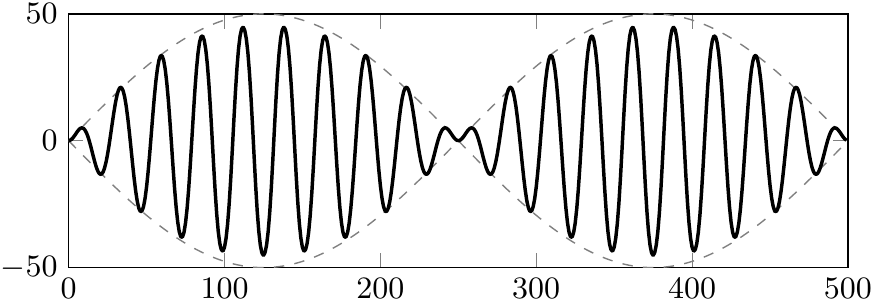}%
\caption{The solid line depicts the output signal of the nonlinear
	amplifier for $B=1$ when stimulated with $s[n]$ in~\eqref{eq:sndef}.}%
\label{fig:inoutcmpb1}%
\end{figure}%
%
\begin{figure}[!t]\centering%
\includegraphics{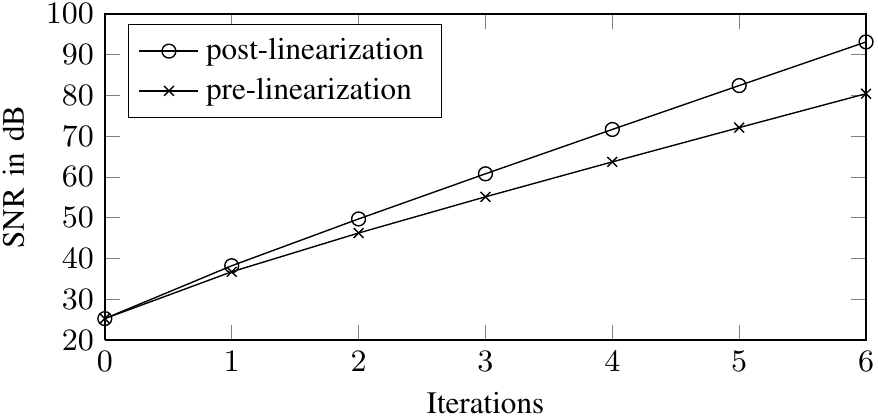}%
\caption{SNR after post- and pre-linearization of the nonlinear amplifier with $B=1$
	with respect to the number of iterations performed in the Richardson equalizer.}%
\label{fig:snrb1}%
\end{figure}

For $B=1$ the output signal of the nonlinear amplifier is moderately
distorted as shown in Fig.~\ref{fig:inoutcmpb1}. In this case, the maxima
of $\spsi{x}{n}$ are at $0.9987$, which is just below one, and thus the
condition in~\eqref{eq:richcondconv} still guarantees convergence.
The linearization performance is depicted in Fig.~\ref{fig:snrb1}.
It can be observed that the improvement in SNR per iteration is still significant, but less compared to
the performance for the mildly distorted signal in Fig.~\ref{fig:snrb075}. This
behavior is a consistent property of the Richardson equalizer, i.e., the
closer the bound imposed by the condition in~\eqref{eq:richcondconv}
is attained, the slower is the convergence.
Another characteristic observable in Fig.~\ref{fig:snrb1} is the deterioration in performance of
the pre-linearizer compared to the post-linearizer. Although this is, in part, explained by the
argument provided in the previous section, another issue becomes
evident here. In particular, the pre-linearizer operates on the signal $y[n]=s[n]$
and convergence is ensured for $x[n]$ bounded by~\eqref{eq:simprexinf}.
Thus, it is implicitly assumed that the maximum gain of the pre-linearizer is
one. For mild distortions this is approximately true, but for moderate and severe distortions
the pre-linearizer needs to compensate the saturation effect by amplification of the
input signal. Consequently, some samples are outside the bound of guaranteed convergence
and deteriorate the performance, a case which is investigated in more detail in the next section.

\subsection{Severe Distortion}

%
\begin{figure}[!t]\centering%
\includegraphics{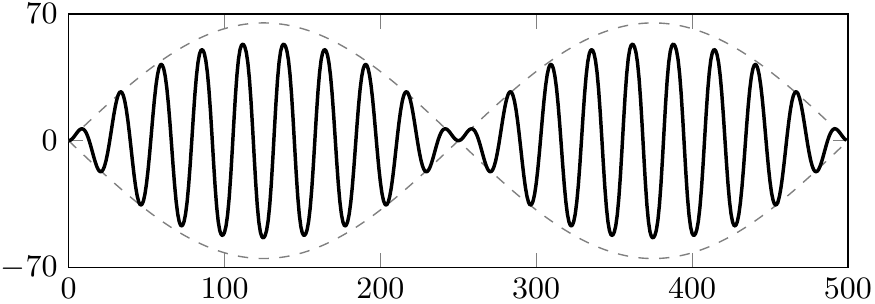}%
\caption{The solid line depicts the output signal of the nonlinear
	amplifier for $B=1.3$ when stimulated with $s[n]$ in~\eqref{eq:sndef}.}%
\label{fig:inoutcmpb130}%
\end{figure}%
%
\begin{figure}[!t]\centering%
\includegraphics{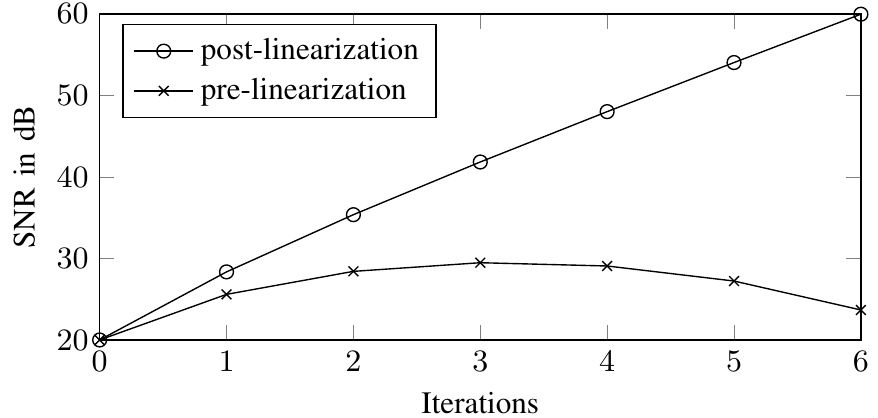}%
\caption{SNR after post- and pre-linearization of the nonlinear amplifier with $B=1.3$
	with respect to the number of iterations performed in the Richardson equalizer.}%
\label{fig:snrb130}%
\end{figure}
%
\begin{figure}[!t]\centering%
\subfloat[Error without linearization.]{%
\includegraphics{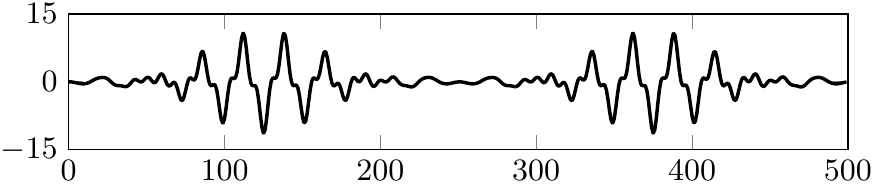}}\hfil%
\subfloat[Error after pre-linearization with $6$ iterations.]{%
\includegraphics{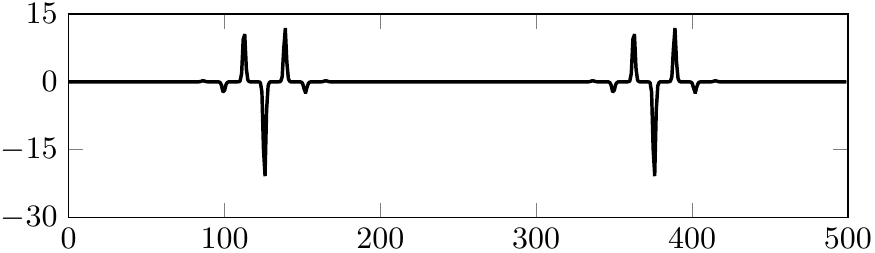}}%
\caption{Error reduction for pre-linearization of the nonlinear amplifier with $B=1.3$.
	In (b), it can be observed that for the majority of samples the iteration has
	practically converged. However, at the six major signal peaks the iteration
	starts to diverge, causing the SNR to deteriorate.}%
\label{fig:errorreductionpreb130}%
\end{figure}%

For the moderate distortion discussed in the previous section, the bound
imposed by the condition in~\eqref{eq:richcondconv} is nearly attained.
Therefore, it represents the amount of distortion for which convergence
is guaranteed by this condition. However, \eqref{eq:richcondconv} is derived
by the repeated application of the triangle inequality, utilization of the
supremum norm as an upper bound on individual samples, and the upper bound
in~\eqref{eq:e0ltx}, cf. Appendix~\ref{ccderi1}. As the latter bound is not
exact and the worst case in terms of the other bounds appears to be quite
improbable, it is reasonable to try to linearize more severly distorted signals.
To this end, consider the input of the nonlinear amplifier to be bounded by $B=1.3$.
The corresponding output signal is depicted in Fig.~\ref{fig:inoutcmpb130}.
In this case, $\spsi{x}{n}$ is between $1.6788$ and $1.7808$, which is
significantly above one, and thus the condition in~\eqref{eq:richcondconv}
cannot guarantee convergence.
However, the linearization performance in Fig.~\ref{fig:snrb130} illustrates that
the post-linearizer still converges. In case of the pre-linearizer, the issue identified
in the previous section becomes more severe. Due to the strong saturation, the pre-linearizer
needs to substantially amplify the signal peaks. The SNR initially improves because the
iteration converges for the majority of samples, but finally it starts to deteriorate
because of the divergence at the signal peaks as illustrated in Fig.~\ref{fig:errorreductionpreb130}.
In this context, it is important to recognize that due to the structure of the Richardson
equalizer in~\eqref{eq:richeq} and the memory in $H_n$ the divergence can propagate to
neighboring samples with repeated iterations.
Concluding, the condition for convergence is rather conservative and the linearization
method presented in this paper may be utilized in cases of more severe distortion. However,
it should be kept in mind that the rate of convergence decreases and that
it may involve the risk of divergence induced by signal peaks.

\subsection{Modeling Errors}

\begin{figure}[!t]\centering%
\includegraphics{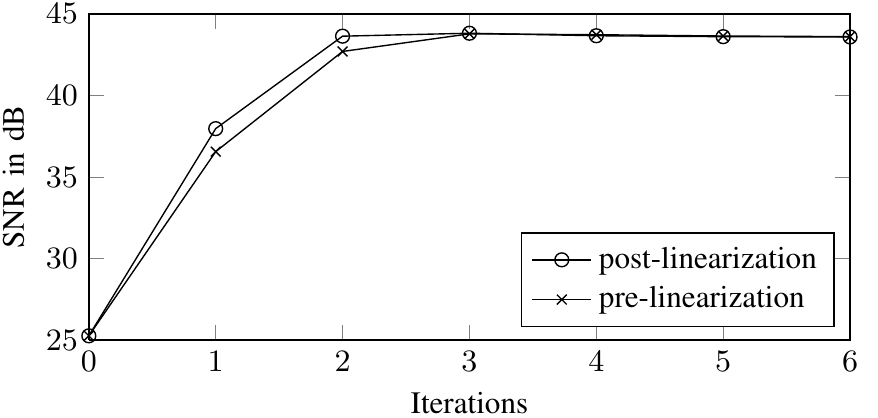}%
\caption{SNR after post- and pre-linearization of the nonlinear amplifier with $B=1$
	in the presence of modeling errors.}%
\label{fig:snrb1:moderr}%
\end{figure}

The previous examples assumed that the nonlinear amplifier is perfectly
known. However, in practice the model is usually only an
approximation of the actual nonlinear system and, therefore, the impact of
modeling errors on the linearization performance is of interest.
In the following, this is investigated by employing the erroneous coefficient vectors
\begin{align*}
	\vm{c}_1 &= ( 0.99,  0.025,  0.03)\\
	\vm{c}_3 &= (-0.37, -0.1,   -0.01)\\
	\vm{c}_5 &= (-0.29, -0.03)
\end{align*}
in the Volterra system used for linearization, which represents
a significant modeling error with respect to the nonlinear amplifier
based on the coefficient vectors in~\eqref{eq:coeffnlamp}. The
corresponding linearization performance is depicted in Fig.~\ref{fig:snrb1:moderr}
for $B=1$. For the augmented Volterra system with modeling errors, the maxima of
$\spsi{x}{n}$ are at $0.9809$ and, therefore, convergence is guaranteed.
Indeed, the SNR increases in the first two iterations but, subsequently,
the convergence stalls. This stems from the
fact that the linearizers effectively linearize a \emph{different} Volterra
system, i.e., they converge to a different solution and the linearization
performance is limited by this deviation. Consequently, the
proposed linearization method is robust against modeling errors if the
condition for convergence is satisfied and the limitation in linearization
performance is determined by the severity of the modeling errors.

%
%

\section{Conclusion}
	\label{conclusion}

In this paper, a novel real-time capable method for the linearization of nonlinear systems
modeled by a time-varying discrete-time Volterra series was presented.
To this end, an alternative view on the Volterra series was established,
which resembles the description of an LTV system. Based on this system model,
a systematic approach to the modification of certain linear iterative methods was
proposed that permits their use for linearization. The modification was
presented for the Richardson iteration and its
utilization for post- and pre-linearization was discussed in detail.
It was shown that the resulting method is a generalization of the equalizer for
linear weakly time-varying systems in~\cite{soudan2012} and that it may
be regarded as a generalization of the post-linearizer in~\cite{nowak1997}
and the $P$th-order inverse.
Due to the iterative structure of the proposed linearizers, their computational cost
scales with the required accuracy via the employed number of iterations.
With the presentation of a simply verifiable
condition for convergence, a practical tool to determine the applicability
of the method was established. By means of the linearization of a time-varying
nonlinear amplifier, the application of the proposed method was exemplified
and properties thereof were discussed. It was shown that the condition for
convergence can guarantee the applicability for mildly to moderately
distorted signals. In this case, the linearizers perform very well and the
iteration converges fast. Consequently, one or two iterations of the
underlying fixed-point iteration may already suffice to achieve a
practically relevant accuracy. It was demonstrated that the
method is also applicable to severely distorted signals, however,
by trading slower convergence and the risk of divergence.
Finally, it was shown that the method
is robust against modeling errors and that the performance penalty
is determined by the severity of the modeling errors.

The proposed method offers considerable potential for future research. Specifically,
the method was presented on the basis of the Richardson iteration, but it is
not limited to this particular linear iterative method. Therefore, other linear
iterative methods like the Jacobi or Gau\ss-Seidel iteration may be explored as
well, which includes the derivation of the corresponding modified iteration and
condition for convergence. Additionally, a preconditioner in terms of a
relaxation parameter might be incorporated to improve the convergence behavior.
In specific scenarios, where further information about the input
signal is available, a more elaborate performance analysis might be performed
by means of the derivation of a worst-case and average rate of convergence.
These results may also aid the design of practical systems as they support
the choice of the employed number of iterations.

\appendices

\section{Convergence of a Time-Varying\\ Discrete-Time Volterra Series}
	\label{bibo}

A time-varying discrete-time Volterra series $H_n$ is \emph{convergent}
if the output of the system is finite for a given input signal~\cite{boyd1984}. For a further
analysis of convergence, let the supremum norm $\norminf{\vm{x}}$
be the bound on the input signal $x[n]$. Using the triangle inequality, the bound
$\norminf{\vm{x}}$ on the input, and $\normone{h_{p,n}}$ in~\eqref{eq:hpnnorm}, it follows
from $y[n]$ in~\eqref{eq:vsdef} that
\begin{align}
	\abs{y[n]} &= \bigg|\sum_{p=1}^\infty \sum_{k_1,\dots,k_p\in\Z}
	              h_{p,n}[k_1,\dots,k_p]\prod_{i=1}^p x[n-k_i]\bigg| \notag\\
	           &\leq \sum_{p=1}^\infty \sum_{k_1,\dots,k_p\in\Z}
	              \abs{h_{p,n}[k_1,\dots,k_p]}\prod_{i=1}^p \abs{x[n-k_i]} \notag\\
	           &\leq \sum_{p=1}^\infty \normone{h_{p,n}}\cdot\norminf{\vm{x}}^p \;.
	\label{eq:yub}
\end{align}
Let the \emph{bound function} $f_n(\norminf{\vm{x}})$ at time instant $n$ be defined as (cf.~\cite{boyd1984})
\begin{align}
	f_n(\norminf{\vm{x}}) = \sum_{p=1}^\infty \normone{h_{p,n}}\cdot\norminf{\vm{x}}^p\;.
	\label{eq:bndfct}
\end{align}
Then it follows from~\eqref{eq:yub} that
\begin{align*}
	\norminf{\vm{y}} = \sup_{n\in\Z} \abs{y[n]}
	                 \leq \sup_{n\in\Z} f_n(\norminf{\vm{x}})\;.
\end{align*}
Consequently, if a time-varying discrete-time Volterra series $H_n$
satisfies the condition
\begin{align}
	\sup_{n\in\Z} f_n(\norminf{\vm{x}}) < \infty
	\label{eq:gbfltinfty}
\end{align}
it converges for all input signals bounded by $\norminf{\vm{x}}$.
The bound function $f_n(\norminf{\vm{x}})$ in~\eqref{eq:bndfct} is a power series with non-negative coefficients
and, therefore, is finite for $\norminf{\vm{x}}<R_n$, where the radius of
convergence $R_n$ is given by~\cite{rudin1964,boyd1984}
\begin{align*}
	R_n = \Big[ \limsup_{p\rightarrow\infty} \normone{h_{p,n}}^{1/p} \Big]^{-1}\;.
\end{align*}
This implies that a time-varying discrete-time Volterra series $H_n$
satisfies~\eqref{eq:gbfltinfty} and, thus, converges if 
\begin{align*}
	\norminf{\vm{x}} < R = \inf_{n\in\Z} R_n
\end{align*}
in which $R$ is the \emph{radius of convergence}.

\section{Lipschitz Continuity of a Time-Varying Discrete-Time Volterra Series}
	\label{errprop}

A time-varying discrete-time Volterra series $H_n$ is \emph{Lipschitz continuous} if
\begin{align}
	\norminf{ \vm{A}_{\vm{x}^{(r)}}\vm{x}^{(r)} - \vm{A}_{\vm{x}}\vm{x} } \leq
		\kappa\cdot \norminf{ \vm{x}^{(r)} - \vm{x} }
	\label{eq:lipdef}
\end{align}
holds, where the system model in Section~\ref{model} is used, $\kappa$ is non-negative
and finite, and $\vm{x}$ and $\vm{x}^{(r)}$ are two input signal vectors
with the corresponding coefficient matrices $\vm{A}_{\vm{x}}$ and
$\vm{A}_{\vm{x}^{(r)}}$ in~\eqref{eq:axdef} and~\eqref{eq:axrdef}, respectively.
In the following, it is shown that a \emph{convergent} time-varying discrete-time
Volterra series $H_n$ is Lipschitz continuous.\footnote{In fact, 
for a time-varying discrete-time Volterra series $H_n$ with a radius of convergence
$R>0$, $H_n$ stimulated by the input vector $\vm{x}$ is continuous if $\norminf{\vm{x}} < R$
and Lipschitz continuous if $\norminf{\vm{x}} < R'<R$, cf. the proof
for the continuous-time Volterra series in~\cite{boyd1984}.}
The approach below is an adaptation of the corresponding proof for
the time-invariant continuous-time Volterra series in~\cite{boyd1984}.
Let $\vm{x}$ and $\vm{x}^{(r)}$ be two input vectors, where the difference
is given by $\vm{e}^{(r)}$ in~\eqref{eq:er1def}. Furthermore, let
\begin{align}
	\norminf{\vm{x}} + \norminf{\vm{e}^{(r)}} < R
	\label{eq:xerltr}
\end{align}
in which $R$ is the radius of convergence of $H_n$. Thus, $H_n$ is convergent
for $\vm{x}$ and $\vm{x}^{(r)}$ because $\norminf{\vm{e}^{(r)}}\geq 0$ and
\begin{align}
	\norminf{\vm{x}^{(r)}} \leq \norminf{\vm{x}} + \norminf{\vm{e}^{(r)}}
	\label{eq:xrleqxer}
\end{align}
respectively, where the latter is obtained from the definition of $\vm{e}^{(r)}$
in~\eqref{eq:er1def} by taking the supremum norm
and applying the triangle inequality. Using~\eqref{eq:axdef}, \eqref{eq:axrdef},
$\sgamma{x}{n}^{(1)}[p,k_1,\ldots,k_p]$ defined in~\eqref{eq:gammadef} in Appendix~\ref{gammaub},
and the triangle inequality, the upper bound
\begin{flalign*}
	&\norminf{ \vm{A}_{\vm{x}^{(r)}}\vm{x}^{(r)} - \vm{A}_{\vm{x}}\vm{x} } &\\
		&\;\ = \sup_{n\in\Z} \bigg|
				\sum_{p=1}^\infty \sum_{k_1,\dots,k_p\in\Z} h_{p,n}[k_1,\dots,k_p]
				\sgamma{x}{n}^{(1)}[p,k_1,\ldots,k_p] \bigg| \\
		&\;\ \leq \sup_{n\in\Z}
				\sum_{p=1}^\infty \sum_{k_1,\dots,k_p\in\Z} \abs{h_{p,n}[k_1,\dots,k_p]}\cdot
				\abs{\sgamma{x}{n}^{(1)}[p,k_1,\ldots,k_p]}
\end{flalign*}
is obtained. Using the upper bound~\eqref{eq:gammaabsub2} in Appendix~\ref{gammaub} on
$\abs{\sgamma{x}{n}^{(1)}[p,k_1,\ldots,k_p]}$ as well as $\normone{h_{p,n}}$
in~\eqref{eq:hpnnorm} and the bound function $f_n(\norminf{\vm{x}})$ in~\eqref{eq:bndfct} enables
\begin{flalign}
	&\norminf{ \vm{A}_{\vm{x}^{(r)}}\vm{x}^{(r)} - \vm{A}_{\vm{x}}\vm{x} } &\notag\\
		&\;\ \leq \sup_{n\in\Z}
				\sum_{p=1}^\infty \normone{h_{p,n}}\cdot
				\Big[ (\norminf{\vm{x}} + \norminf{\vm{e}^{(r)}})^{p} - \norminf{\vm{x}}^{p} \Big] \notag\\
		&\;\ = \sup_{n\in\Z} \Big[ f_n(\norminf{\vm{x}} + \norminf{\vm{e}^{(r)}}) - f_n(\norminf{\vm{x}}) \Big]\;.
	\label{eq:axryrubgbf}
\end{flalign}
From the mean value theorem it follows that~\cite{rudin1964,boyd1984}
\begin{align}
	f_n(\norminf{\vm{x}} + \norminf{\vm{e}^{(r)}}) - f_n(\norminf{\vm{x}})
		= f'_n(\zeta)\cdot\norminf{\vm{e}^{(r)}}
	\label{eq:mvtgbf}
\end{align}
where $f'_n$ is the derivative of $f_n$ and
\begin{align*}
	\norminf{\vm{x}} \leq \zeta \leq \norminf{\vm{x}} + \norminf{\vm{e}^{(r)}}\;.
\end{align*}
Using this relation in~\eqref{eq:axryrubgbf} yields
\begin{align*}
	&\norminf{ \vm{A}_{\vm{x}^{(r)}}\vm{x}^{(r)} - \vm{A}_{\vm{x}}\vm{x} }
		\leq \norminf{\vm{e}^{(r)}} \sup_{n\in\Z} f'_n(\zeta)
\end{align*}
which corresponds to~\eqref{eq:lipdef} where
\begin{align}
	\kappa = \sup_{n\in\Z} f'_n(\zeta)\;.
	\label{eq:kappadef}
\end{align}
Due to~\eqref{eq:gbfltinfty},~\eqref{eq:xerltr}, and~\eqref{eq:mvtgbf}, $\kappa$ in~\eqref{eq:kappadef}
is indeed non-negative and finite, which completes the proof.

\section{Condition for Convergence for\\ the Richardson Equalizer}
	\label{ccderi1}

In this appendix, it is proven that~\eqref{eq:richcondconv}
guarantees convergence of the Richardson equalizer in~\eqref{eq:richeq}
with the initialization in~\eqref{eq:richinit} by showing that
it is a \emph{sufficient} condition for~\eqref{eq:ermondec} to hold.

\subsection{Problem Statement}

Using $\vm{e}^{(r+1)}$ in~\eqref{eq:er1eq} and the definition of
$\vm{A}_{\vm{x}}$ and $\vm{A}_{\vm{x}^{(r)}}$ in~\eqref{eq:axdef}
and~\eqref{eq:axrdef}, respectively, the error
$e[n]^{(r+1)}$ in iteration $r+1$ at time instant $n$ can be expressed as
\begin{align*}
	e[n]^{(r+1)}
		&= \sum_{k_1\in\Z} (\delta[k_1]-\sg{x}{n}[k_1]) e[n-k_1]^{(r)} \\
				&\qquad + \sum_{k_1\in\Z} (\sg{x^{(r)}}{n}[k_1] - \sg{x}{n}[k_1]) x[n-k_1]^{(r)}\;.
\end{align*}
Therewith, the supremum norm of $\vm{e}^{(r+1)}$ is upper bounded using
the triangle inequality as
\begin{align*}
	\norminf{\vm{e}^{(r+1)}}
		&= \sup_{n\in\Z} \abs{e[n]^{(r+1)}} \\
		&= \sup_{n\in\Z} \bigg| \sum_{k_1\in\Z} (\delta[k_1]-\sg{x}{n}[k_1]) e[n-k_1]^{(r)} \\
				&\qquad + \sum_{k_1\in\Z} (\sg{x^{(r)}}{n}[k_1] - \sg{x}{n}[k_1]) x[n-k_1]^{(r)} \bigg| \\
		&\leq \sup_{n\in\Z} \left[
			\abs{\salpha{x}{n}(\vm{e}^{(r)})} + \abs{\sbeta{x}{n}(\vm{e}^{(r)})} \right]
\end{align*}
where the first and second sum is given by
$\salpha{x}{n}(\vm{e}^{(r)})$ and $\sbeta{x}{n}(\vm{e}^{(r)})$
in~\eqref{eq:alphadef} and~\eqref{eq:betadef} in
Appendix~\ref{alphaub} and~\ref{betaub}, respectively. Using the upper bounds~\eqref{eq:alphaub}
and~\eqref{eq:betaub} for $\abs{\salpha{x}{n}(\vm{e}^{(r)})}$ and $\abs{\sbeta{x}{n}(\vm{e}^{(r)})}$
derived in Appendix~\ref{alphaub} and~\ref{betaub}, respectively, it follows that
\begin{align}
	\norminf{\vm{e}^{(r+1)}}
    \leq&\ \norminf{\vm{e}^{(r)}} \sup_{n\in\Z} \seta{x}{n}(\norminf{\vm{e}^{(r)}})
	\label{eq:er1leqereta}
\end{align}
where
\begin{align}
	\begin{split}
		\seta{x}{n}(\norminf{\vm{e}^{(r)}}) &= \sum_{k_1\in\Z} \abs{\delta[k_1]-h_{1,n}[k_1]} \\
					&\qquad\ + \sum_{p=2}^\infty \normone{h_{p,n}} \cdot \swtilde{x}(p,\norminf{\vm{e}^{(r)}})
	\end{split}
	\label{eq:etardef}
\end{align}
and
\begin{align}
	\begin{split}
	&\swtilde{x}(p,\norminf{\vm{e}^{(r)}}) = \norminf{\vm{x}}^{p-1} \\
		&\quad\qquad+ \norminf{\vm{x}^{(r)}}
				\sum_{l=1}^{p-1} \binom{p-1}{l} \norminf{\vm{x}}^{p-1-l}\norminf{\vm{e}^{(r)}}^{l-1}\;.
	\end{split}
	\label{eq:wtildedef}
\end{align}
It can be observed that $\seta{x}{n}(\norminf{\vm{e}^{(r)}})$ is monotonically
increasing with respect to the non-negative argument $\norminf{\vm{e}^{(r)}}$,
which is relevant later on. Indeed, only
$\swtilde{x}(p,\norminf{\vm{e}^{(r)}})$ depends on $\norminf{\vm{e}^{(r)}}$
and, as can be seen in~\eqref{eq:wtildedef}, it is a polynomial of degree $p-2$ with
non-negative coefficients and, therefore, monotonically increasing.
According to~\eqref{eq:er1leqereta}, if
\begin{align}
	\sup_{n\in\Z} \seta{x}{n}(\norminf{\vm{e}^{(r)}}) < 1
	\label{eq:condeta}
\end{align}
holds for all iterations $r\geq 0$,
the condition for convergence in~\eqref{eq:ermondec} holds as well. In the
following, sufficient conditions for~\eqref{eq:condeta} to hold in the
first iteration are derived. Subsequently,
this result is used for an inductive proof of convergence under the same conditions.

\subsection{Error Reduction in First Iteration}
	\label{ccderi1:base}
	
Due to the initialization in~\eqref{eq:richinit} and the system model in~\eqref{eq:mtxeq},
the initial error $\vm{e}^{(0)}$ is given by
\begin{align*}
	\vm{e}^{(0)} = \vm{x} - \vm{x}^{(0)}
	             = \vm{x} - \vm{y}
	             = (\vm{I} - \vm{A}_{\vm{x}})\vm{x}\;.
\end{align*}
With the definition of $\vm{A}_{\vm{x}}$ in~\eqref{eq:axdef}, the supremum norm
of the initial error can be identified as
\begin{align*}
	\norminf{\vm{e}^{(0)}} = \sup_{n\in\Z} \abs{\salpha{x}{n}(\vm{x})}\;.
\end{align*}
Due to the upper bound on $\abs{\salpha{x}{n}(\vm{x})}$ in~\eqref{eq:alphaub}
in Appendix~\ref{alphaub}, this norm is upper bounded by
\begin{align*}
	\norminf{\vm{e}^{(0)}} \leq
	\norminf{\vm{x}} \sup_{n\in\Z} \setatilde{x}{n}
\end{align*}
where
\begin{align*}
	\setatilde{x}{n} = \sum_{k_1\in\Z} \abs{\delta[k_1]-h_{1,n}[k_1]}
			+ \sum_{p=2}^\infty \normone{h_{p,n}} \cdot \norminf{\vm{x}}^{p-1}\;.
\end{align*}
A comparison of $\setatilde{x}{n}$ to $\seta{x}{n}(\norminf{\vm{e}^{(r)}})$
in~\eqref{eq:etardef} reveals that $\setatilde{x}{n}\leq \seta{x}{n}(\norminf{\vm{e}^{(r)}})$
for all $\norminf{\vm{e}^{(r)}}\geq 0$. Consequently, any condition that ensures~\eqref{eq:condeta}
enforces
\begin{align*}
	\sup_{n\in\Z} \setatilde{x}{n} < 1
\end{align*}
as well. Therefore, it can be assumed that the initial error is bounded by
\begin{align}
	\norminf{\vm{e}^{(0)}} < \norminf{\vm{x}}
	\label{eq:e0ltx}
\end{align}
because a contradiction in this inequality would also invalidate~\eqref{eq:condeta}.
Due to the fact that $\seta{x}{n}(\norminf{\vm{e}^{(r)}})$ in~\eqref{eq:etardef}
is a monotonically increasing function for non-negative arguments, it follows
that
\begin{align}
	\seta{x}{n}(\norminf{\vm{e}^{(0)}}) \leq \seta{x}{n}(\norminf{\vm{x}})\;.
	\label{eq:etaxne0ltetaxnx}
\end{align}
Consequently, requiring
\begin{align}
	\sup_{n\in\Z} \seta{x}{n}(\norminf{\vm{x}}) < 1
	\label{eq:supetaxnlt1}
\end{align}
ensures that~\eqref{eq:condeta} holds for the first
iteration and, therefore, $\norm{\vm{e}^{(1)}} < \norm{\vm{e}^{(0)}}$.
Rewriting~\eqref{eq:er1def}, taking the supremum norm, and applying the
triangle inequality leads to~\eqref{eq:xrleqxer} and permits the bound
\begin{align*}
	\norminf{\vm{x}^{(0)}}
		\leq \norminf{\vm{x}} + \norminf{\vm{e}^{(0)}}
		< 2 \norminf{\vm{x}}\;.
\end{align*}
Using this upper bound in~\eqref{eq:wtildedef} for the argument $\norminf{\vm{x}}$
gives\footnote{A comparison of $\sum_{l=1}^{p-1} \binom{p-1}{l}$ to the binomial theorem
shows that it corresponds to $2^{p-1}-1$.}
\begin{align*}
	\swtilde{x}(p,\norminf{\vm{x}})
		&< \norminf{\vm{x}}^{p-1}\left[ 1 + 2 \sum_{l=1}^{p-1} \binom{p-1}{l} \right] \\
		&= \norminf{\vm{x}}^{p-1}(2^p-1) \;.
\end{align*}
Finally, utilizing this upper bound on $\swtilde{x}(p,\norminf{\vm{x}})$ in~\eqref{eq:etardef} to
obtain an upper bound on $\seta{x}{n}(\norminf{\vm{x}})$ and, subsequently, using the result
in~\eqref{eq:supetaxnlt1} leads to the condition for convergence in~\eqref{eq:richcondconv}.

\subsection{Inductive Proof of Convergence}

Convergence of the Richardson equalizer can be ensured by induction if
\begin{align}
	\sup_{n\in\Z} \seta{x}{n}(\norminf{\vm{e}^{(r+1)}}) \leq \sup_{n\in\Z} \seta{x}{n}(\norminf{\vm{e}^{(r)}})
	\label{eq:etar1leqetar}
\end{align}
holds, as, due to~\eqref{eq:er1leqereta}, this implies that~\eqref{eq:ermondec} holds.
The condition for convergence in~\eqref{eq:richcondconv} establishes the basis
\begin{align*}
	\sup_{n\in\Z} \seta{x}{n}(\norminf{\vm{e}^{(0)}}) < 1
\end{align*}
which follows from \eqref{eq:etaxne0ltetaxnx} and~\eqref{eq:supetaxnlt1}. As a
consequence of~\eqref{eq:er1leqereta}, this basis implies $\norm{\vm{e}^{(1)}} < \norm{\vm{e}^{(0)}}$.
As $\seta{x}{n}(\norminf{\vm{e}^{(r)}})$ is a monotonically increasing function for
non-negative arguments, it follows that~\eqref{eq:etar1leqetar} holds for $r=0$
and, due to~\eqref{eq:er1leqereta},~\eqref{eq:ermondec} holds for $r=1$.
This induction step can be repeated ad infinitum and, therefore, completes the proof.

\section{Upper Bound for $\abs{\salpha{x}{n}(\vm{e}^{(r)})}$}
	\label{alphaub}

In this appendix, an upper bound for the absolute value of
\begin{align}
	\salpha{x}{n}(\vm{e}^{(r)})
		= \sum_{k_1\in\Z} (\delta[k_1]-\sg{x}{n}[k_1]) e[n-k_1]^{(r)}
	\label{eq:alphadef}
\end{align}
is derived. Using the triangle inequality and the supremum norm $\norminf{\vm{e}^{(r)}}$
as an upper bound on $e[n-k_1]^{(r)}$ yields
\begin{align*}
	\abs{\salpha{x}{n}(\vm{e}^{(r)})}
		&\leq \sum_{k_1\in\Z} \abs{\delta[k_1]-\sg{x}{n}[k_1]}\cdot\abs{e[n-k_1]^{(r)}} \\
		&\leq \norminf{\vm{e}^{(r)}} \sum_{k_1\in\Z} \abs{\delta[k_1]-\sg{x}{n}[k_1]}\;.
\end{align*}
Substitution of $\sg{x}{n}[k_1]$ with~\eqref{eq:gxn} and application of the
triangle inequality permits the upper bound
\begin{flalign*}
	&\abs{\salpha{x}{n}(\vm{e}^{(r)})}
		\leq \norminf{\vm{e}^{(r)}} \bigg[ \sum_{k_1\in\Z} \abs{\delta[k_1]-h_{1,n}[k_1]} \\
			&\quad\quad\quad\quad+ \sum_{p=2}^\infty \sum_{k_1,\ldots,k_p\in\Z}
				\abs{h_{p,n}[k_1,\ldots,k_p]}\prod_{i=2}^p \abs{x[n-k_i]}\bigg]\;.
\end{flalign*}
With the supremum norm $\norminf{\vm{x}}$ as an upper bound on $x[n-k_i]$
and $\normone{h_{p,n}}$ in~\eqref{eq:hpnnorm}, $\abs{\salpha{x}{n}(\vm{e}^{(r)})}$
is upper bounded by
\begin{align}
	\begin{split}
	\abs{\salpha{x}{n}(\vm{e}^{(r)})}
		&\leq \norminf{\vm{e}^{(r)}} \bigg[ \sum_{k_1\in\Z} \abs{\delta[k_1]-h_{1,n}[k_1]} \\
			&\quad\quad\quad\quad\quad\quad
			+ \sum_{p=2}^\infty \normone{h_{p,n}} \cdot \norminf{\vm{x}}^{p-1} \bigg]\;.
	\end{split}
	\label{eq:alphaub}
\end{align}

\section{Upper Bound for $\abs{\sbeta{x}{n}(\vm{e}^{(r)})}$}
	\label{betaub}

In this appendix, an upper bound for the absolute value of
\begin{align}
	\sbeta{x}{n}(\vm{e}^{(r)})
		= \sum_{k_1\in\Z} (\sg{x^{(r)}}{n}[k_1] - \sg{x}{n}[k_1]) x[n-k_1]^{(r)}
	\label{eq:betadef}
\end{align}
is derived. Using the triangle inequality and the supremum norm $\norminf{\vm{x}^{(r)}}$
as an upper bound on $x[n-k_1]^{(r)}$ yields
\begin{align}
	\abs{\sbeta{x}{n}(\vm{e}^{(r)})}
		&\leq \sum_{k_1\in\Z} \abs{\sg{x^{(r)}}{n}[k_1] - \sg{x}{n}[k_1]}\cdot\abs{x[n-k_1]^{(r)}} \notag\\
		&\leq \norminf{\vm{x}^{(r)}} \sum_{k_1\in\Z} \abs{\sg{x^{(r)}}{n}[k_1] - \sg{x}{n}[k_1]}\;.
	\label{eq:betaabssumub}
\end{align}
Substituting the impulse responses with~\eqref{eq:gxn} and~\eqref{eq:gxrn}, respectively,
applying the triangle inequality, and utilizing $\sgamma{x}{n}^{(2)}[p,k_2,\ldots,k_p]$
defined in~\eqref{eq:gammadef} in Appendix~\ref{gammaub} results in the upper bound
\begin{flalign*}
	&\sum_{k_1\in\Z} \abs{\sg{x^{(r)}}{n}[k_1] - \sg{x}{n}[k_1]} &\\
		&= \sum_{k_1\in\Z} \bigg| \sum_{p=2}^\infty \sum_{k_2,\ldots,k_p\in\Z}
			h_{p,n}[k_1,\ldots,k_p] \sgamma{x}{n}^{(2)}[p,k_2,\ldots,k_p] \bigg| \\
		&\leq \sum_{p=2}^\infty \sum_{k_1,\ldots,k_p\in\Z}
			\abs{h_{p,n}[k_1,\ldots,k_p]}\cdot\abs{\sgamma{x}{n}^{(2)}[p,k_2,\ldots,k_p]}\;.
\end{flalign*}
Using the upper bound~\eqref{eq:gammaabsub1} on
$\abs{\sgamma{x}{n}^{(2)}[p,k_2,\ldots,k_p]}$ in Appendix~\ref{gammaub} 
and $\normone{h_{p,n}}$ in~\eqref{eq:hpnnorm} yields
\begin{align}
	&\hspace{-2.9mm}\sum_{k_1\in\Z} \abs{\sg{x^{(r)}}{n}[k_1] - \sg{x}{n}[k_1]} \notag\\
		&\hspace{-2.9mm}\leq \norminf{\vm{e}^{(r)}} \sum_{p=2}^\infty \normone{h_{p,n}}
					\sum_{l=1}^{p-1} \binom{p-1}{l} \norminf{\vm{x}}^{p-1-l}\norminf{\vm{e}^{(r)}}^{l-1}.
	\label{eq:gxrgxabssumub}
\end{align}
Finally, using~\eqref{eq:gxrgxabssumub} in~\eqref{eq:betaabssumub} permits the upper bound
\begin{align}
\begin{split}
	\abs{\sbeta{x}{n}(\vm{e}^{(r)})}
		&\leq \norminf{\vm{e}^{(r)}} \sum_{p=2}^\infty \normone{h_{p,n}}\cdot \norminf{\vm{x}^{(r)}} \\
		&\qquad\times \sum_{l=1}^{p-1} \binom{p-1}{l} \norminf{\vm{x}}^{p-1-l}\norminf{\vm{e}^{(r)}}^{l-1}\;.
	\label{eq:betaub}
\end{split}
\end{align}

\section{Upper Bound for $\abs{\sgamma{x}{n}^{(q)}[p,k_q,\ldots,k_p]}$}
	\label{gammaub}

In this appendix, an upper bound for the absolute value of
\begin{align}
	\sgamma{x}{n}^{(q)}[p,k_q,\ldots,k_p] =
		\prod_{i=q}^p x[n-k_i]^{(r)} - \prod_{i=q}^p x[n-k_i]
	\label{eq:gammadef}
\end{align}
is derived, where $1\leq q\leq p$. Using the definition of $\vm{e}^{(r)}$
in~\eqref{eq:er1def}, the absolute value of $\sgamma{x}{n}^{(q)}[p,k_q,\ldots,k_p]$
can be expressed as
\begin{align}
\begin{split}
	&\abs{\sgamma{x}{n}^{(q)}[p,k_q,\ldots,k_p]} \\
		&\quad\ = \bigg| \prod_{i=q}^p (x[n-k_i] - e[n-k_i]^{(r)}) - \prod_{i=q}^p x[n-k_i] \bigg|\;.
\end{split}
	\label{eq:gammaxer}
\end{align}
If the first product therein is expanded, it contains a summand
that cancels with the second product. In order to find
an upper bound on the remaining terms, the first product is analyzed.
Using the triangle inequality and $\norminf{\vm{x}}$ and $\norminf{\vm{e}^{(r)}}$
as upper bounds on $x[n-k_i]$ and $e[n-k_i]^{(r)}$, respectively, enables
\begin{align*}
	&\bigg| \prod_{i=q}^p (x[n-k_i] - e[n-k_i]^{(r)}) \bigg| \\
		&\qquad\qquad\qquad\qquad \leq \prod_{i=q}^p (\abs{x[n-k_i]} + \abs{e[n-k_i]^{(r)}}) \\
		&\qquad\qquad\qquad\qquad \leq (\norminf{\vm{x}} + \norminf{\vm{e}^{(r)}})^{p-q+1}\;.
\end{align*}
For this bound, the binomial theorem gives
\begin{align*}
	(\norminf{\vm{x}} + \norminf{\vm{e}^{(r)}})^N
		= \norminf{\vm{x}}^N
		+ \sum_{l=1}^{N} \binom{N}{l} \norminf{\vm{x}}^{N-l}\norminf{\vm{e}^{(r)}}^l
\end{align*}
in which $N=p-q+1$. It can be recognized that $\norminf{\vm{x}}^N$ corresponds to the
upper bound of the term that cancels with the second product in~\eqref{eq:gammaxer}
and, therefore,
\begin{align}
\begin{split}
	\abs{\sgamma{x}{n}^{(q)}[p,k_q&,\ldots,k_p]} \\
		&\leq \sum_{l=1}^{p-q+1} \binom{p-q+1}{l} \norminf{\vm{x}}^{p-q+1-l}\norminf{\vm{e}^{(r)}}^l\;.
\end{split}
	\label{eq:gammaabsub1}
\end{align}
Equivalently, this bound can be stated as
\begin{align}
\begin{split}
	\abs{\sgamma{x}{n}^{(q)}[p,k_q,&\ldots,k_p]} \\
		&\leq (\norminf{\vm{x}} + \norminf{\vm{e}^{(r)}})^{p-q+1} - \norminf{\vm{x}}^{p-q+1}\;.
\end{split}
	\label{eq:gammaabsub2}
\end{align}

\section{Kernels of the Volterra System $H_n$ for\\ Post- and Pre-Linearization}
	\label{linkernels}

\subsubsection{Post-Linearization}

For post-linearization, the Volterra system $H_n$ is given by~\eqref{eq:hnbreve}.
From Fig.~\ref{fig:postlinaltsys} and the definition of the Volterra system
in~\eqref{eq:vsdef} it follows that
\begin{align*}
	y[n]
		= \sum_{p=1}^\infty \sum_{\nu_1,\dots,\nu_p\in\Z} \hat{h}_{p,n}[\nu_1,\dots,\nu_p]\prod_{i=1}^p u[n-\nu_i]\;.
\end{align*}
Using the definition of $u[n]$ in~\eqref{eq:linvdef} and $L^{-1}$ in~\eqref{eq:linvdefconv}
results in
\begin{align*}
	y[n]
		&= \sum_{p=1}^\infty \sum_{\nu_1,\dots,\nu_p\in\Z} \hat{h}_{p,n}[\nu_1,\dots,\nu_p] \\
				&\qquad\qquad\qquad\qquad\qquad\quad\times
				\prod_{i=1}^p \sum_{l\in\Z} q[l] x[n-\nu_i-l] \\
		&= \sum_{p=1}^\infty \sum_{\nu_1,\dots,\nu_p\in\Z} \sum_{l_1,\dots,l_p\in\Z}
				\hat{h}_{p,n}[\nu_1,\dots,\nu_p] \\
				&\qquad\qquad\qquad\qquad\qquad\quad\times
				\prod_{i=1}^p q[l_i] x[n-\nu_i-l_i]\;.
\end{align*}
The substitution $k_i=\nu_i+l_i$, for $i=1,\ldots,p$, and
partition of the product yields
\begin{align*}
	y[n]
		&= \sum_{p=1}^\infty \sum_{k_1,\dots,k_p\in\Z} \bigg[
				\sum_{l_1,\dots,l_p\in\Z}  \hat{h}_{p,n}[k_1-l_1,\dots,k_p-l_p] \\
				&\qquad\qquad\qquad\qquad\qquad\quad\times
				\prod_{j=1}^p q[l_j] \bigg] \prod_{i=1}^p x[n-k_i]\;.
\end{align*}
A comparison to~\eqref{eq:vsdef} shows that $y[n]$ is given by~\eqref{eq:vsdef}
with the Volterra kernels in~\eqref{eq:hpnpostlindef}.

\subsubsection{Pre-Linearization}

For pre-linearization, the Volterra system $H_n$ is given by~\eqref{eq:hnring}.
From Fig.~\ref{fig:prelinaltsys} and the definition of $L^{-1}$ in~\eqref{eq:linvdefconv}
it follows that
\begin{align}
	y[n] = \sum_{l\in\Z} q[l] v[n-l]\;.
	\label{eq:prekerny}
\end{align}
Furthermore, from Fig.~\ref{fig:prelinaltsys} and the definition of the Volterra
system in~\eqref{eq:vsdef} it follows that $v[n]$ is given by
\begin{align}
	v[n]
		= \sum_{p=1}^\infty \sum_{\nu_1,\dots,\nu_p\in\Z} \check{h}_{p,n}[\nu_1,\dots,\nu_p]\prod_{i=1}^p x[n-\nu_i]\;.
	\label{eq:prekernv}
\end{align}
Using~\eqref{eq:prekernv} in~\eqref{eq:prekerny} yields
\begin{align*}
	y[n] &= \sum_{p=1}^\infty \sum_{\nu_1,\dots,\nu_p\in\Z} \sum_{l\in\Z}
				\check{h}_{p,n-l}[\nu_1,\dots,\nu_p] q[l] \\
				&\qquad\qquad\qquad\qquad\qquad\qquad\quad\times
				\prod_{i=1}^p x[n-l-\nu_i]\;.
\end{align*}
The substitution $k_i=l+\nu_i$, for $i=1,\ldots,p$, results in
\begin{align*}
	y[n]
		&= \sum_{p=1}^\infty \sum_{k_1,\dots,k_p\in\Z} \bigg[ \sum_{l\in\Z}
				\check{h}_{p,n-l}[k_1-l,\dots,k_p-l] q[l] \bigg] \\
				&\qquad\qquad\qquad\qquad\qquad\qquad\quad\times
				\prod_{i=1}^p x[n-k_i]\;.
\end{align*}
A comparison to~\eqref{eq:vsdef} shows that $y[n]$ is given by~\eqref{eq:vsdef}
with the Volterra kernels in~\eqref{eq:hpnprelindef}.

\bibliographystyle{IEEEtran}
\bibliography{tsp2014_hotz_vogel}

\vspace*{3em}

\end{document}